\documentclass[useAMS,usenatbib]{mn2e}
\usepackage[normalem]{ulem}

\usepackage{tabularx,ragged2e,booktabs}
\usepackage{natbib}
\usepackage{graphicx}
\usepackage{amsmath,amsfonts,amssymb}
\usepackage{multirow}
\usepackage{hyperref}
\usepackage{color}
\usepackage{cases}
\usepackage[figure]{hypcap}

\hypersetup{colorlinks, citecolor=blue, pdfduplex=DuplexFlipLongEdge}
\hypersetup{pdftitle=nonthermal paper, pdfauthor=Andrew Chael, pdfsubject=Astrophysics}
\def\sgra{{Sgr~A$^{\ast} $} }
\def\sgraa{{Sgr~A$^{\ast} $}}
\def\adi{\Gamma_\text{int}}
\def\d{\partial}

\title[Nonthermal Electrons in Accretion Simulations]{Evolving Nonthermal Electrons in Simulations of Black Hole Accretion}
\author[A. Chael et al.]
{Andrew A. Chael$^1$\thanks{\hbox{E-mail: achael@cfa.harvard.edu}},
Ramesh Narayan$^1$,
Aleksander S{\c a}dowski$^{2,3}$
\\
$^1$Harvard-Smithsonian Center for Astrophysics, 60 Garden Street, Cambridge, MA 
02138, USA\\
$^2$MIT Kavli Insitute for Astrophysics and Space Research, 77 Massachusetts 
Ave, Cambridge, MA 02139, USA \\
$^3$Akuna Capital, 585 Massachusetts Ave, Cambridge, MA 02139, USA
}

\begin{document}
\maketitle

\begin{abstract}
Current simulations of hot accretion flows around black holes assume
either a single-temperature gas or, at best, a two-temperature gas
with thermal ions and electrons.  However, processes like magnetic
reconnection and shocks can accelerate electrons into a nonthermal
distribution, which will not quickly thermalise at the very
low densities found in many systems. Such nonthermal electrons have been invoked 
to explain the infrared and X-ray spectra and
strong variability of Sagittarius A* (\sgraa), the black hole at the
Galactic Center. We present a method for self-consistent evolution of
a nonthermal electron population in the GRMHD code \texttt{KORAL}. The
electron distribution is tracked across Lorentz
factor space and is evolved in space and time, in parallel with
thermal electrons, thermal ions, and radiation. In the
present study, for simplicity, energy injection into the nonthermal distribution is
taken as a fixed fraction of the local electron viscous heating rate. 
Numerical results are presented for a model with a low mass
accretion rate similar to that of \sgraa. We find that the presence of
a nonthermal population of electrons has negligible effect on the
overall dynamics of the system. Due to our simple uniform particle injection prescription, 
the radiative power in the nonthermal simulation is enhanced at large radii. 
The energy distribution of the nonthermal electrons shows a synchrotron cooling
break, with the break Lorentz factor varying with location and time,
reflecting the complex interplay between the local viscous heating rate,
magnetic field strength, and fluid velocity.
\end{abstract}


\section{Introduction}
\label{sec::intro}

Nearly every galaxy is thought to host a supermassive black hole at
its centre, which accretes gas and liberates a large fraction of its
binding energy in the form of radiation and outflows. Black hole
accretion discs in active galactic nuclei often have luminosities
close to the Eddington limit and are among
the most luminous steady sources in the universe.  In contrast, the
black hole source at the centre of the Milky Way, Sagittarius A*
(\sgraa), has
a low luminosity $\sim 10^{-9}$ of Eddington \citep{Falcke98, Genzel03, 
Baganoff03}, and a correspondingly low mass accretion rate $\lesssim 10^{-7}$ of 
Eddington \citep{Agol2000, Bower2003, Marrone_2007}. The low luminosity of Sgr 
A* is due both to its low accretion rate and the low radiative efficiency of the 
accreting gas \citep{Ichimaru77, Rees1982, NarayanYi_a, NarayanYi_b, 
Blandford99, Narayan98, Quat99b}. Such radiatively inefficient or 
advection-dominated accretion flows (ADAFs) are geometrically thick and optically thin, 
and the gas is extremely hot (see \citealt{Yuan14} for a review).

When the density of the accreting gas is low, as in systems accreting
below about $10^{-3}$ of the Eddington rate, Coulomb collisions
between electrons and ions become rare, and the electron-ion
thermalisation time exceeds the accretion time. Electrons and ions can
then have different temperatures, with the ratio set by the balance
between the viscous heating rates of the two species, the rate of
energy transfer from ions to electrons by Coulomb coupling, and the
rate of radiative cooling of the electrons.  At the lowest densities,
when the electron-electron collision time-scale becomes
sufficiently long, collisions will not completely relax the electron
distribution function to a local Maxwellian \citep{Maha97}. Even if
the bulk of the electron distribution function is thermal, processes
like shocks and magnetic reconnection \citep{Sironi11, Sironi14} can
accelerate a small fraction of the electrons into a relativistic
nonthermal distribution, which will persist
for a long time because of the lack of collisions.

While traditional ADAF models emitting via thermal synchrotron radiation can 
describe the bulk of the emission from \sgra in the `submm bump' around 
$10^{12}$ Hz \citep{NarayanYi_b,Narayan98, Quat99a}, the quiescent infrared and 
unresolved X-ray emission in the spectrum are most easily explained with hybrid 
models that include a small population of high-energy electrons in a power-law 
tail \citep{Ozel2000,Yuan2003,Yuan2004,Brod_Loeb}.  In addition, \sgra is highly 
variable from the millimeter band to X-rays \citep{Genzel03, Eckart06, Dodds11}. 
Analysis of 3D general relativistic magnetohydrodynamic (GRMHD) simulations has shown that synchrotron emission from the 
turbulent plasma, lensed by the central black hole, can produce variability in 
the infrared and sub-mm \citep{Chan_15b}. Furthermore, thermal synchrotron emission in the infrared can be
inverse Compton upscattered by the thermal electrons to higher frequencies, 
producing correlated variability in the  X-rays \citep{Ressler17}.

Observations of strong correlated 
flares ($\gtrsim 10$ times the quiescent flux) in the X-ray and infrared suggest that,
in addition to the usual thermal electrons, there is rapid and localized 
injection of a broad nonthermal, power-law electron distribution that 
radiates strongly via synchrotron emission \citep{Yusef06, Dodds09, Neilsen}. 
Recently, \citet{Ball_16} took 3D GRMHD simulations from \citet{Narayan2012} and 
\citet{Sadowski2013} and moved a small percentage of the radiating electrons to a high-energy 
power-law distribution in regions of high magnetization.  They showed that such 
a localized injection of nonthermal electrons produces correlated, broad-band 
variability in the infrared and X-ray flux, with variability properties similar 
to those observed in \sgraa.

Numerical simulations of \sgra and other low accretion rate systems
have typically been done within the context of single-fluid ideal
magnetohydrodynamics (e.g. \citealt{Hawley00, Tchekhovskoy10, McKinney12, 
Narayan2012}),
though recently \citet{Chandra15, Foucart, Chandra17} have
incorporated the non-ideal MHD effect of anisotropic heat
conduction. In order to produce spectra and images from simulations while
taking into account the weak coupling between electrons and ions, the
temperature ratio of the electrons to the total gas is usually set
manually in post-processing. This ratio can be fixed gobally,
(e.g. \citealt{Moscibrodzka_09, Dexter_10}), vary depending on the
region of the flow (disc or jet) under consideration
(e.g. \citealt{Moscibrodzka_14, Chan_15b}), or computed as a function of fluid
properties (e.g. \citealt{Shch_12}).  Including radiation from
nonthermal electrons is less common, but some recent investigations
have added a population of electrons with a power-law distribution of
Lorentz factor during post-processing to investigate the effect on the
quiescent spectrum and sub-mm image size of Sgr A* \citep{Mao2016} and its
variability properties \citep{Ball_16}.

Recent work (\citealt{Ressler15,KORAL16,Ressler17}) has extended single-fluid 
GRMHD simulations by introducing methods for self-consistent evolution of 
separate ion and electron populations. \citet{Ressler15,Ressler17} evolve a 
single-temperature non-radiative accretion flow and then evolve viscously heated 
electrons during a separate post-processing step. \citet{KORAL16} evolve ions 
and electrons simultaneously as two fluid components in a radiative magnetohydrodynamic simulation (GRRMHD), 
self-consistently including the effects of Coulomb coupling and radiative cooling
(though these are not dynamically important, at least in the case of Sgr A*).  
Although the ions and electrons are treated as separate fluids in these studies, 
each species is still assumed to be independently in thermal equilibrium.

In this paper, we take the next logical step. Namely, in addition to
evolving thermal ions, thermal electrons and radiation in a GRRMHD
simulation, as in \citet{KORAL16}, we also evolve a population of
nonthermal electrons.  The exchange of energy and momentum among the
various fluid populations and the radiation field is accounted for at
each time step during the evolution. The nonthermal electrons are
heated by a prescribed fraction of the total viscous heating rate and
they further gain and lose energy by adiabatic compression and
expansion, Coulomb coupling, inverse Compton scattering, and radiative
cooling. The algorithm is implemented in the GRRMHD code
\texttt{KORAL} \citep{KORAL13, KORAL14, KORAL16}.

In Section~\ref{sec::phys}, we review the standard GRMHD equations and
present the additional equations needed for evolving the nonthermal
population of electrons in the presence of radiation and Coulomb
coupling. In Section~\ref{sec::num}, we describe the numerical
algorithm, and in Section~\ref{sec::test} we discuss a number of
simple test problems that we have used to validate the code. In
Section~\ref{sec::sgra}, we present initial results of a simulation of
a \sgraa-like accretion flow, which includes nonthermal electrons. We
conclude with a summary in Section~\ref{sec::summary}.

\section{Physics}
\label{sec::phys}

\subsection{Fluid Populations}
\label{sec::fluidpops}

GRMHD simulations typically track a single magnetized perfect fluid as a 
function of position and time. The single fluid is described by the gas density 
$\rho$, internal energy density $u$, and four-velocity $u^\mu$. The pressure $p$ 
is related to the internal energy $u$ via the adiabatic index $\adi$,
\begin{equation}
 p=(\adi-1)u.
\end{equation}
In the ideal MHD approximation, the fluid frame electric field vanishes due to 
the high plasma conductivity. As a result, the electromagnetic field can be 
specified entirely via a magnetic field four-vector $b^\mu$ \citep{Gammie03}.
The MHD stress-energy tensor $T^\mu_{\;\;\;\nu}$ then takes the form
\begin{equation}
 \label{eq::tmunu}
 T^\mu_{\;\;\nu} = \left(\rho + u + p + b^2\right)u^\mu u_\nu + \left(p + 
\frac{1}{2}b^2\right)\delta^\mu_{\;\;\;\nu} - b^\mu b_\nu.
\end{equation}

In the present work, the fluid consists of three populations: thermal ions, 
thermal electrons, and an isotropic distribution of nonthermal electrons. We 
assume that all three populations move with the same velocity $u^\mu$. This assumption automatically
preserves local charge neutrality and simplifies the evolution equations for the nonthermal
spectrum (Section ~\ref{sec::nthev}). Under this approximation, equation~\eqref{eq::tmunu} 
remains a valid description of the total stress-energy, although the equation of state relating $p$ and $u$ changes as 
explained below.

The electrons contribute negligibly to the mass density, hence,
\begin{equation}
\label{eq::rho}
\rho = \mu_i m_p n_i,
\end{equation}
where $m_p$ is the proton mass, $\mu_i$ is the ion mean molecular weight, and 
$n_i$ is the fluid frame number density of ions. Denoting the electron mean 
molecular weight by $\mu_e$, charge neutrality enforces the constraint
\begin{equation}
\mu_e(n_{e\,\text{th}} + n_{e\, \text{nth}}) = \mu_i n_i = \rho/m_p,
\end{equation}
where $n_{e\,\text{th}},$ and $n_{e\, \text{nth}}$ are the number densities of 
the thermal and nonthermal electrons, respectively. In the simulations presented 
in this paper we consider only pure ionized hydrogen: $\mu_i=\mu_e=1$. 


All three fluid populations can have substantial contributions to the net energy 
density and pressure,
\begin{align}
u &= u_i + u_{e\, \text{th}} + u_{e\, \text{nth}}, \nonumber \\
p &= p_i + p_{e\,\text{th}} + p_{e\, \text{nth}}.
\end{align}
The energy densities and pressures of the thermal species are determined by 
their respective temperatures $T_{i,e}$ and corresponding adiabatic indices 
$\Gamma_{i,e}$,
\begin{align}
\label{eq::perfectgas}
p_{i,e\,\text{th}} &= n_{i,e\,\text{th}}k_BT_{i,e}, \\
u_{i,e\,\text{th}} &= \frac{p_{i,e\,\text{th}}}{\Gamma_{i,e}(\theta_{i,e})-1}. 
\end{align}
For each species, the adiabatic index $\Gamma(\theta)$ is a function of temperature through the 
dimensionless ratio $\theta = k_B T/m c^2$, transitioning from 
$\Gamma=5/3$ for non-relativistic particles ($\theta \ll 1$) to $\Gamma=4/3$ for 
relativistic particles ($\theta \gg 1$). 

Instead of directly tracking the temperatures or energy densities of
the individual species, we work with the electron and ion entropy per particle
$s_{i,e}$, which allows us to break up the evolution into adiabatic
and non-adiabatic steps. For a non-degenerate relativistic gas, there
exist exact closed form expressions for the adiabatic index
$\Gamma(\theta)$ and the entropy per particle $s(\theta, n)$ 
\citep{Chandra39}. However, because the exact expressions involve
computationally expensive Bessel functions and are not easy to invert, 
we use approximate forms. Our approach is based on a fitting function to the specific
heat at constant volume, which we can integrate to find expressions
for the internal energy, (see Appendix A of \citealt{KORAL16}) \footnote{\citealt{KORAL16} derived 
equation~\eqref{eq::adiabspecies} in Appendix A, but used a simpler 
fitting function for $u(\theta)$ to enable direct inversion to solve for 
$\theta$. Here we have chosen to use equation~\eqref{eq::adiabspecies} in order 
to maintain consistency in our approximations, though at a small additional computational 
cost. On the occasions that we need to solve for $\theta$ from $u$, we use a 
Newton-Raphson solver to invert equation~\eqref{eq::adiabspecies}, which 
converges rapidly.}
\begin{equation}
 \label{eq::adiabspecies}
 \frac{u(\theta)}{p(\theta)} = \frac{1}{\Gamma(\theta) - 1} = 3-\frac{3}{5\theta}\ln\left[1+\frac{5\theta}{2}\right],
\end{equation}
and the entropy per particle,
\begin{equation}
\label{eq::entpp}
s = k_B\ln\left[\frac{\theta^{3/2}(\theta+2/5)^{3/2}}{n}\right].
\end{equation}

We assume that nonthermal particles are isotropic in the fluid rest
frame, with a distribution $n(\gamma)$ in Lorentz factor $\gamma$. In
the current work, we consider $n(\gamma)$ over a range of $\gamma$
from a minimum $\gamma_{\rm{min}}$ to a maximum
$\gamma_{\rm{max}}$. The number density, energy density, and pressure
of the nonthermal electrons are then simply given by integrals over
the distribution $n(\gamma)$,
\begin{align}
\label{eq::nthint}
n_{e\,\text{nth}} &= \int_{\gamma_{\rm{min}}}^{\gamma_{\rm{max}}} n(\gamma)\,\rm{d}\gamma,  \\
\label{eq::uthint}
u_{e\,\text{nth}} &= m_e c^2\int_{\gamma_{\rm{min}}}^{\gamma_{\rm{max}}}n(\gamma)(\gamma-1)\,\rm{d}\gamma,  \\
\label{eq::nthints}
p_{e\,\text{nth}} &= \frac{m_e c^2}{3}\int_{\gamma_{\rm{min}}}^{\gamma_{\rm{max}}} n(\gamma)(\gamma-\gamma^{-1})\,\rm{d}\gamma,
\end{align}
where $m_e$ is the electron mass.\footnote{The $m_ec^2(\gamma-\gamma^{-1})/3$ 
factor in the integrand for the pressure $p_{e\,\text{nth}}$ is just $p^2/3E$, 
where $p^2 = m_ec^2(\gamma^2-1)$ is the square of the particle momentum, 
$E=\gamma m_ec^2$ is the particle energy, and the $1/E$ factor comes from the 
relativistically invariant measure $d^3\mathbf{p}/E$ (e.g. \citealt{Mihalas}).}

When necessary, we calculate the net adiabatic index of the combined 
three-species  fluid directly,
\begin{equation}
\label{eq::adiabindex}
\adi = 1+\frac{p}{u} = 1 + \frac{p_i + p_{e\,\text{th}} + p_{e\, 
\text{nth}}}{u_i + u_{e\, \text{th}} + u_{e\, \text{nth}}},
\end{equation}
using equations~(\ref{eq::perfectgas}---\ref{eq::adiabspecies}) for the thermal 
quantities and equations~\eqref{eq::uthint} and \eqref{eq::nthints} for the 
nonthermal energy and pressure. 

\subsection{Radiation}
\label{sec::radiation}

In addition to the three fluid components described above, we
concurrently evolve an independent fluid to represent radiation. We
specify the radiation field using the M1 closure scheme
\citep{Levermore84}, as described in \citet{KORAL13,KORAL14,McKinney14}. In 
effect we assume that, at each spacetime
point, there exists a `radiation frame' in which the radiation is
isotropic. Thus the frequency-integrated radiation field is described
by its energy density $\bar{E}$ in the radiation frame and the
timelike four velocity $u^\mu_r \neq u^\mu$ of this frame
(the evolution equations which determine the radiation  frame 
four-velocity are described in Section~\ref{sec::phot_num}).
In an arbitrary frame, the radiation stress energy tensor then takes the form
\begin{equation}
 R^\mu_{\;\;\nu} = \frac{4}{3}\bar{E}_ru^\mu_r u_{r\nu} + 
\frac{1}{3}\bar{E}_r\delta^\mu_{\;\;\nu}.
\end{equation}
Throughout this work, quantities in the radiation rest frame are
denoted with bars, and quantities in the fluid frame are denoted with
hats. In particular, while $\bar{E}$ is the radiation energy density
evolved by the code, the fluid frame quantity $\hat{E}$ is what enters
into the equations describing the interactions between radiation and
the fluid components.

We also independently evolve the radiation frame photon number $\bar{n}_r$, 
which is useful for calculating the radiation temperature. Under the assumption 
that the radiation spectrum is a grey body \citep{Compt15}, the radiation 
temperature in the fluid frame is 
\begin{equation}
 \label{eq::rtemp}
 T_r = \frac{\hat{E}}{2.7012 \, k_B \hat{n}_r},
\end{equation}
where $\hat{E}$ and $\hat{n}_r$ are the radiation energy density and photon 
number transformed to the fluid frame. 

\subsection{GRRMHD equations}
\label{sec::grrmhd}

The conservation equations that govern the evolution of the fluid, magnetic 
field, and radiation field are
\begin{align}
\label{eq::GRRMHD}
(\rho u^\mu)_{;\mu} & = 0, \\
T^{\mu}_{\;\;\nu;\mu} &= G_\nu, \\
R^{\mu}_{\;\;\nu;\mu} &= -G_\nu, \\
\label{eq::GRRMHD4}
F^{*\mu}_{\;\;\;\;\nu;\mu} &=0.
\end{align}

Here, $F^{*\mu\nu} = b^\mu u^\nu - b^\nu u^\mu$ is the dual of the MHD Maxwell tensor, and  
$G_\nu$ is the four-force density that couples the evolution of the 
radiation and gas (see equation~\ref{eq::rad_flux} and \ref{eq::rad_flux_2}).
We assume that both the thermal and nonthermal electrons radiate isotropically 
in the fluid rest frame. Hence, the total energy-momentum flux from radiation to 
the gas is
\begin{align}
\label{eq::rad_flux}
\hat{G}^0 &= \tilde{\rho}\,(\kappa_{P,a}\hat{E} - 4\pi\kappa_{P,e}\hat{B}) + 
\hat{G}^0_{IC \, \text{th}} + \hat{G}^0_\text{nth} ,   \\
\label{eq::rad_flux_2}
\hat{G}^i &= (\tilde{\rho}\kappa_{R} + \rho\kappa_{es})\hat{F}^i.
\end{align}
In these equations, the $\kappa$ factors are the total, frequency-averaged 
opacities for the thermal radiative processes; the distinctions between the 
different factors are explained below. $\hat{G}^0_{IC \, \text{th}}$ is the 
thermal energy loss from inverse Compton scattering, and $\hat{G}^0_\text{nth}$ 
is the energy loss to radiation from the nonthermal population. $\tilde{\rho}$ 
is the fluid density reduced to account for the nonthermal population (see 
below), $\hat{F}^i$ is the radiation momentum flux, and $\hat{B} = \sigma T_e^4 / 
\pi$ is the electron blackbody radiance.

We ignore absorption by the nonthermal population (Section~\ref{sec::nthev}). 
Therefore, in the rest frame, the nonthermal population only contributes an 
emission factor to the energy term $\hat{G}^0_\text{nth}$. The contribution to the 
radiative power from the nonthermal electrons is the integral of the radiative 
cooling rate over the distribution,
\begin{equation} 
\label{eq::g0nth}
\hat{G}^0_\text{nth} = m_e c^2 \int_{\gamma_{\rm{min}}}^{\gamma_{\rm{max}}} 
n(\gamma) \dot{\gamma}_\text{rad}\,\rm{d}\gamma.
\end{equation}
The quantity $\dot{\gamma}_\text{rad}$ represents the cooling rate of an 
electron with energy $\gamma m_e c^2$ from radiative processes in the fluid rest 
frame; it is always negative. The total radiative cooling rate has contributions 
from synchrotron, bremsstrahlung, and inverse Compton scattering,
\begin{equation}
\label{eq::gdotrad}
 \dot{\gamma}_\text{rad} = 
\dot{\gamma}_\text{syn}+\dot{\gamma}_\text{brem}+\dot{\gamma}_\text{IC},
\end{equation}
where $\dot{\gamma}_\text{syn},\dot{\gamma}_\text{brem},\dot{\gamma}_\text{IC}$ 
are given by equations~(\ref{eq::syncoolrate}, \ref{eq::bremcoolrate}, 
\ref{eq::iccoolrate}), respectively.

The thermal electrons contribute an emission term 
$4\pi\tilde{\rho}\kappa_{P,e}\hat{B}$ and an absorption term 
$\tilde{\rho}\kappa_{P,a}\hat{E}$ to the energy flux. Because we ignore 
absorption by the nonthermal electrons, the momentum flux comes entirely from 
the thermal absorption term $\tilde{\rho}\kappa_{R}\hat{F}^i$ and the electron 
scattering term  $\rho\kappa_{es}\hat{F}^i$, where $\hat{F}^i$ is the radiation 
flux in the fluid frame. 

We use frequency-averaged, grey opacities for thermal synchrotron and 
bremsstrahlung emission. Following the suggestion of \citet{Mihalas}, we use the 
Planck averaged mean opacities $\kappa_{P,e}$ and $\kappa_{P,a}$  weighted for 
emission  and absorption in the energy equation $(\hat{G}^0)$, and we use the Rosseland 
mean opacity $\kappa_R$ in the momentum equation $(\hat{G}^i)$. The full 
expressions for these opacities as a function of number density and 
temperature are given in \citet{KORAL16}. The electron scattering opacity is 
$\kappa_{es}$; it includes a Klein-Nishina factor that lowers the scattering 
cross section at high photon energies \citep{Buchler76},
\begin{equation}
\label{eq::kappaes}
 \rho \kappa_{es} = (n_{e\,\text{th}} + n_{e\, \text{nth}})\sigma_T \left[1 + 
\frac{T_r}{4.5\times10^8 \, \text{K}}\right]^{0.86} \;\;\text{cm}^{-1}.
\end{equation}
Note that the density multiplying most of the thermal synchrotron and 
bremsstrahlung opacities in equations\eqref{eq::rad_flux} and 
\eqref{eq::rad_flux_2} is $\tilde{\rho}$, which corresponds to thermal electrons 
alone,
\begin{equation}
 \tilde{\rho} = \rho\frac{n_\text{e,\,th}}{n_\text{e,\,th}+n_\text{e\,nth}}.
\end{equation}
The full density $\rho$ is used for the electron scattering opacity, since 
the nonthermal electrons also scatter the emission. 

Finally, inverse Compton scattering off of thermal electrons contributes the 
last term, $\hat{G}^0_\text{IC\,\text{th}}$. The full expression for 
$\hat{G}^0_\text{IC\,\text{th}}$ can be found in \citet{Compt15}.
 

\subsection{Nonthermal population evolution equation}
\label{sec::nthev}

The evolution equation for the nonthermal distribution can be derived by taking 
angular moments of the relativistic Boltzmann equation and imposing the 
requirement that the distribution be isotropic in the fluid rest frame 
\citep{Lindquist_1966,Webb_85,Webb_89}. The isotropy assumption truncates the 
hierarchy of moment equations and leaves a single equation,
\begin{align}
\label{eq::elecev}
\left[n(\gamma)u^\alpha\right]_{;\alpha} &= - \frac{\d}{\d 
\gamma}\left[\dot{\gamma}_\text{tot}n(\gamma)\right] + Q_I(\gamma), \\
\label{eq::gdottot}
\dot{\gamma}_\text{tot} &=  \dot{\gamma}_\text{adiab} + \dot{\gamma}_\text{C} + 
\dot{\gamma}_\text{rad}.
\end{align}
Aside from the injection (source) term $Q_I(\gamma)$,
equation~\eqref{eq::elecev} is essentially a conservation equation in
five dimensions: four dimensions correspond to space and time (left-hand 
side of equation~\ref{eq::elecev}), and the fifth dimension
corresponds to the fluid frame particle Lorentz factor $\gamma$,
through which particles move with velocity
$\dot{\gamma}_\text{tot}$. This velocity is broken into three parts:
$\dot{\gamma}_\text{adiab}$ from adiabatic heating and cooling due to
gas compression and expansion, $\dot{\gamma}_\text{C}$ from cooling
due to the (weak) Coulomb coupling with the thermal electrons, and
$\dot{\gamma}_\text{rad}$ from the energy lost to radiation
(equation~\ref{eq::gdotrad}). Since we assume nonthermal electrons
only emit and do not absorb, $\dot{\gamma}_\text{rad}$ is always
negative; furthermore, the Coulomb coupling term
$\dot{\gamma}_\text{C}$ is also negative, since the nonthermal
population by assumption consists of particles more energetic
than the thermal electrons that they couple to.

The adiabatic `cooling' rate $\dot{\gamma}_\text{adiab}$ can be positive or 
negative, depending on whether the gas is compressing or expanding. This term 
can be derived from the relativistic Boltzmann equation without interaction 
terms \citep{Webb_89},
\begin{equation}
\label{eq::gdotadiab}
\dot{\gamma}_\text{adiab} = -\frac{1}{3}u^\alpha_{\;\;;\alpha}(\gamma -\gamma^{-1}).
\end{equation}
It is negative when the gas expands,
$\equiv u^\alpha_{\;\;;\alpha} > 0$, and it is positive when the gas
is compressed.

The term $Q_I(\gamma)$ in equation~\eqref{eq::elecev} is the rate of
injection of high energy electrons from the thermal to the nonthermal
distribution at a given $\gamma$. In principle, $Q_I(\gamma)$ is a
function of local conditions and depends on microscopic plasma
processes that accelerate electrons into the nonthermal
distribution. For simplicity, in this study, we assume that the
electrons are injected with a power-law distribution with a constant
index $p$,
\begin{equation}
 \label{eq::powlawinj}
  Q_I(\gamma) = C \gamma^{-p}.
\end{equation}

In addition, we assume that the total rate of energy injection into the high 
energy population is a fixed fraction $\delta_\text{nth}$ of the total electron 
viscous heating rate $\delta_e q^v$. This determines the normalization $C$ in 
equation~\eqref{eq::powlawinj},
\begin{equation}
 m_ec^2 \int (\gamma -1) Q_I(\gamma) = \delta_\text{nth}\delta_eq^v.
\end{equation}
Thus, given the total viscous heating rate $q^v$, which we compute
numerically from the simulation (see equation~\ref{eq::vischeat}),
we add a fraction $\delta_e$ of the energy to the electrons, of which
a fraction $\delta_\text{nth}$ goes into the nonthermal
population. These heating fractions are adjustable parameters of the
model; in this work, we follow \citet{Ressler15} and \citet{KORAL16} in using 
the fitting formula of \citet{Howes10} to determine
$\delta_e$ as a function of species temperature and plasma
magnetization, and we set $\delta_\text{nth}$ equal to a constant
value ($\delta_\text{nth} = 0.015$ for the nonthermal run described in
Section~\ref{sec::sgra}).

Apart from $\dot{\gamma}_\text{adiab}$, the model includes additional cooling 
rates, $\dot{\gamma}_\text{\,syn}, \, \dot{\gamma}_\text{\,brem}, \, 
\dot{\gamma}_\text{\,IC}, \dot{\gamma}_\text{\,C}$, for synchrotron, 
bremsstrahlung, inverse Compton scattering, and Coulomb coupling. We use 
expressions from \citet{Mano} and \citet{Ginzburg}, valid
in the relativistic limit ($\gamma \gg 1$),
\begin{align}
\dot{\gamma}_\text{\,syn} &= 
-1.292\times10^{-11}\left(\frac{B}{1\,\text{G}}\right)^2 
\,\gamma^2\;\text{s}^{-1}, \label{eq::syncoolrate} \\
\dot{\gamma}_\text{\,brem} &= - 
1.37\times10^{-16}\left(\frac{n_i}{1\,\text{cm}^{-3}}\right)\,\gamma\,
(\ln\gamma+0.36)\;\text{s}^{-1}, \label{eq::bremcoolrate} \\
\dot{\gamma}_\text{\,IC} &= -3.25\times 10^{-8}\left(\frac{\hat{E}}{1 
\,\text{erg}\,\text{cm}^{-3}}\right)\gamma^2 \;F_{KN}(\gamma) \;\text{s}^{-1}, 
\label{eq::iccoolrate} \\
\dot{\gamma}_\text{\,C} &= 
-1.491\times10^{-14}\left(\frac{n_{e\,\text{th}}}{1\,\text{cm}^{-3}}
\right)\times \nonumber \\
&\;\;\;\;\;\;\;\;\;\;\;\;\;\;\;\;\;\;\;\left[\ln\gamma+\ln\left(\frac{n_{e\,
\text{th}}}{1\,\text{cm}^{-3}}\right) + 74.7\right] \;\text{s}^{-1}. 
\label{eq::cccoolrate}
\end{align}

The inverse Compton cooling rate $\dot{\gamma}_\text{IC}$ includes a 
dimensionless Klein-Nishina factor $F_{KN}$ which reduces the cooling rate at 
high $\gamma$. For a thermal distribution of photons at temperature $T_r$, this 
factor is \citep{Mano,Moderski}
\begin{equation}
\label{eq::FKN}
F_{KN}(\gamma) = \left(1+11.2\gamma\frac{kT_r}{m_ec^2}\right)^{-3/2}.
\end{equation}

\subsection{Thermal population evolution}
\label{sec::thev}

The evolution of the thermal ions and electrons is handled as in 
\citet{KORAL16}, with additional terms to describe the new interactions with 
nonthermal electrons. For both species, the thermal entropy per particle evolves according to the 
first law of thermodynamics,
\begin{align}
\label{eq::ent_ev}
T_e(n_{e\,\text{th}}s_eu^\mu)_{;\mu} &= \delta_e(1-\delta_\text{nth}) q^v + 
q_{\text{th}}^C + \hat{G}_\text{th}^0, \\
&\;\;\; + q_{\text{nth}}^C + \left(q^\text{cool} - 
\mu\dot{n}^\text{cool}\right), \nonumber\\
T_i(n_{i}s_iu^\mu)_{;\mu} &= (1-\delta_e) q^v - q_{\text{th}}^C.
\end{align}
The first term on the right-hand side in both equations represents the
viscous heating of the thermal populations.  The total viscous heating rate $q^v$ is identified numerically in our
algorithm, as described later (see equation~\ref{eq::vischeat}). The
fraction of the viscous heating that goes to the thermal ions is $\left(1-\delta_e\right)$, and the 
fraction that goes into thermal electrons is $\left(1-\delta_\text{nth}\right)\delta_e$. The second term
in both equations is the thermal Coulomb coupling $q^C_\text{th}$
between the thermal electron and ion populations; the expression we
use is from \citet{Stepney83}, and can be found in
\citet{KORAL16}. The third term in the electron entropy equation is
the net emission and absorption of energy from radiation,
$\hat{G}^0_\text{th}$ (equation~\ref{eq::rad_flux}).

The nonthermal population modifies the electron entropy evolution through a 
Coulomb coupling term $q^C_\text{nth}$, which is the total energy gained by the 
thermal electrons due to the Coulomb cooling of the high-energy particles,
\begin{equation}
 q^C_\text{nth} = -m_e c^2 \int n(\gamma) \dot{\gamma}_\text{C} \, \rm{d}\gamma. 
\end{equation}

Finally, in order to conserve the total number of electrons, we assume that when 
nonthermal electrons cool below $\gamma_{\rm{min}}$, they are thermalised and 
join the thermal distribution. These cooling electrons join the thermal 
distribution at a rate $\dot{n}^\text{cool}$,  carrying energy density $q^\text{cool}$. 
The expression $\mu\dot{n}^\text{cool}$, where $\mu$ is the chemical potential, 
accounts for the increase in entropy from the increase in particle number 
density. The energy and particle cooling rates from the nonthermal distribution 
to the thermal distribution are simply the flux of energy and particles at the 
boundary $\gamma_{\rm{min}}$,
\begin{align}
\label{eq::edgeesc}
\dot{n}^\text{cool} &=  
-\left[\dot{\gamma}_\text{tot}n(\gamma)\right]_{\gamma_{\rm{min}}}, \nonumber \\
q^\text{cool} &=  
-\left[m_e\dot{\gamma}_\text{tot}(\gamma-1)n(\gamma)\right]_{\gamma_{\rm{min}}}.
\end{align}

Note that during adiabatic compression, there can be a (small) flux out of the 
nonthermal distribution at $\gamma_{\rm{max}}$; we treat this similarly, adding 
back the energy and particle number lost over this edge to the local thermal 
bath. This treatment is unphysical but necessary to  conserve energy among the three 
species in the simulation. Since the total amount of viscous heating 
is not increased by this procedure (see equation~\eqref{eq::vischeat}), this choice
will not increase the temperature of the thermal electron population above what it would
be in a  simulation without any nonthermal electrons. In any case, due to the steep power-law shape
of the injection functions considered in this study (equation~\eqref{eq::powlawinj}), the outward flux 
at $\gamma_{\rm{max}}$ is always extremely small.s

For the chemical potential $\mu$, we use the following expression derived 
from our approximate form of the entropy, equation~\eqref{eq::entpp},
\begin{align}
\label{eq::muapprox}
 \mu = m_ec^2&\left[1-\frac{3}{5}\ln\left(1+\frac{5}{2}\theta_e\right)\right.  
\nonumber\\ 
 &+\theta_e\left. 
\left(4-\frac{3}{2}\ln\left(\theta_e^2+\frac{2}{5}\theta_e\right)+\ln n_\text{e\;th}\right)\right].
\end{align}

\subsection{Radiation evolution}
\label{sec::phot_num}

The evolution of the radiation energy density $\bar{E}$ and the
radiation frame velocity $u_r$ is determined by the coupled GRRMHD
equations~\eqref{eq::GRRMHD}--\eqref{eq::GRRMHD4}. Evolving the photon
number density requires a separate equation \citep{KORAL15},
\begin{equation}
\label{eq::photev}
(\bar{n}_r u^\mu_{r})_{;\mu} = \hat{\dot{n}}_r ,
\end{equation}
where $\hat{\dot{n}}_r$ is the frame-invariant photon production rate 
\citep{Compt15},
\begin{equation}
\label{eq::nrdot}
\hat{\dot{n}}_r = \hat{\dot{n}}_{r, \, \text{syn}} + 
\hat{\dot{n}}_\text{brem\,,th} + \hat{\dot{n}}_\text{brem\,,nth} - 
\tilde{\rho}\kappa_{n,a} \hat{n}_r.
\end{equation}
The first term in equation~\eqref{eq::nrdot} is from synchrotron
emission of both thermal and nonthermal electrons (the number of
photons emitted in synchrotron is independent of the energy of the
emitting particle),
\begin{equation}
\hat{\dot{n}}_{r, \, 
\text{syn}}=1.46\times10^5\left(\frac{B}{1\,\text{G}}\right)(n_{e\,\text{th}}+n_
{e\,\text{n}}).
\end{equation}
The second term is the production of photons from thermal
bremsstrahlung emission (see \citealt{KORAL16}). The third term gives
the corresponding rate of photon emission by bremsstrahlung from the
nonthermal distribution. For an electron at $\gamma$, we approximate
the bremsstrahlung photon production by assuming that photons are only
produced with energy $h\nu = \gamma m_ec^2$,
\begin{equation}
\hat{\dot{n}}_\text{brem\,,nth} = \int_{\gamma_{\rm{min}}}^{\gamma_{\rm{max}}} 
\frac{\dot{\gamma}_\text{brem}}{\gamma} \, n(\gamma)\,\rm{d}\gamma.
\end{equation}
Finally, the last term is the photon loss rate from absorption by the thermal 
electrons, which can be written in terms of a number absorption opacity 
$\kappa_{n,a}$ (see \citealt{KORAL16}).

\section{Numerical Methods}
\label{sec::num}

We have implemented the equations in Section~\ref{sec::phys} in the
GRRMHD code \texttt{KORAL} \citep{KORAL13, KORAL14, KORAL16}. The
nonthermal electron distribution $n(\gamma)$ is sampled in $N$ equally
spaced logarithmic bins over a range $[\gamma_{\rm{min}},
  \gamma_{\rm{max}}].$ These quantities $n(\gamma_j)$ are $N$
additional primitive quantities which we evolve in parallel with the
remaining GRRMHD and thermodynamic primitives. The full vector of
primitives $P_i$ consists of the fluid density $\rho$, energy density
$u$, fluid velocity $u^i$; the magnetic field $B^i$, radiation energy
density $\bar{E}$, radiation frame velocity $u_r^i$; the photon
number $n_r$, thermal electron and ion entropy densities
$s_en_\text{e,\,th}$ and $s_in_i$; and the populations $n(\gamma_j)$
of the nonthermal electrons:
\begin{equation}
 \label{eq::prim}
 P = [\rho, u, u^i, B^i, \bar{E}, u_r^i, n_r, s_e n_\text{e,\,th}, s_e 
n_\text{i}, n(\gamma_j)],
\end{equation}
where the index $j$ runs over the $N$ bins sampled in $\gamma$-space. The 
corresponding conserved quantities are
\begin{align}
 \label{eq::cons}
 U &= [ \rho u^0, T^0_0 + \rho u^0, T^0_i, B^i, R^0_0, R^0_i, n_ru^0, 
\nonumber\\
      & \;\;\;\;\;\;\;\;\;s_e n_\text{e,\,th} u^0, s_e n_\text{i} u^0, 
n(\gamma_j) u^0].
\end{align}
The code uses a Newton-Raphson solver to convert from the conserved quantities 
to primitives \citep{KORAL13, KORAL14}. Since the fluid velocity $u^\mu$ is 
uniquely specified by inverting the MHD conserved quantities, to recover 
$n(\gamma_j)$ we simply divide the conserved quantity $n(\gamma_j) u^0$ by $u^0$ 
after the Newton-Raphson solver has found a solution for the MHD and radiation 
primitives.

Fixed floors and ceilings are applied on the evolved quantities as in 
\citet{KORAL13,KORAL14,KORAL16}. We impose an absolute floor on the nonthermal 
distribution $n(\gamma_j)>0$. This is especially necessary when beginning from 
$n(\gamma_j)=0$, as numerical effects can occasionally make $q^v$ negative and 
bring the nonthermal number values below zero. We also impose fixed ceilings to 
prevent the nonthermal number and energy densities from exceeding 50 per cent of the 
total.

\texttt{KORAL} uses a second-order Runge-Kutta scheme to advance the fluid 
quantities in each time step. Within each Runge-Kutta step, there are three main 
substeps: explicit fluid evolution (Section~\ref{sec::explicitev}), nonthermal 
adiabatic evolution and viscous heating (Section~\ref{sec::vischeatev}), and 
implicit radiation and Coulomb coupling (Section~\ref{sec::implicitev}).

\subsection{Explicit fluid evolution}
\label{sec::explicitev}

In the explicit substep, the covariant conservation equations are evolved 
without source terms. 
 
The equations evolved are the GRRMHD equations,
equations~(\ref{eq::GRRMHD}--\ref{eq::GRRMHD4}), the photon number
equation~\eqref{eq::photev}, the thermal entropy
equation~\eqref{eq::ent_ev}, and the nonthermal advection
equation~\eqref{eq::elecev}, all with their right hand sides set to
zero. In particular, we treat the nonthermal distribution at each
point in $\gamma$-space independently and evolve these variables with
the spatial fluid flow. The explicit evolution uses a Lax-Friedrichs
method with van-Leer flux limiters to calculate fluxes of the
conserved quantities at cell faces. Geometrical terms (i.e. the covariant derivative terms
involving Christoffel symbols) are added as source terms at cell
centres. The full explicit advective algorithm is described in \citet{KORAL13, 
KORAL14}.

Physically, entropy density is not exactly conserved on a finite grid, despite 
the form of equation~\eqref{eq::ent_ev}. When we bring two (same-species) finite 
gas parcels together and mix them to an equilibrium energy and temperature, the 
total energy density is constant, but entropy increases. If we solve the 
source-free version of equation~\eqref{eq::ent_ev} with finite-volume methods, 
the entropy will be preserved exactly, and the final energy density will be 
underestimated, causing the viscous heating identified in the next step to be 
systematically too large. 

To avoid this problem, we adopt the simple solution from \citet{KORAL16}
of mixing the entropies from neighboring cells at constant
density.\footnote{As noted there, mixing at constant pressure may be a more 
consistent procedure.} In the explicit evolution of the
adiabatic entropy equation, we 
identify the initial values of the entropy flux on each of the cell walls,
as well as the mixing fractions which they contribute to the total
entropy increase in the cell over the time step. We then take these same
mixing fractions and uses them to instead add up the energy densities
computed from the boundary entropy fluxes, keeping the fluid density fixed. 
Once we have the species energy increase, we invert (using
equations~\ref{eq::adiabspecies} and \ref{eq::entpp}) to find the
final value of the entropy. This approach is computationally
convenient and limits excess viscous heating.
 
\subsection{Adiabatic Nonthermal Evolution and Viscous Heating}
\label{sec::vischeatev}

After evolving the bulk fluid quantities explicitly, we evolve the 
nonthermal distribution in each cell adiabatically through $\gamma$-space to 
provide the appropriate heating or cooling from gas compression or expansion. 
Then we calculate the total energy dissipated and apply it to the thermal and 
nonthermal species using our viscous heating prescription. The steps are as 
follows:

\begin{enumerate}
  \item The nonthermal distribution is evolved under adiabatic 
compression/expansion using the cooling rate $\dot{\gamma}_\text{adiab}$ in 
equation~\eqref{eq::gdotadiab}. From equation~\eqref{eq::elecev}, after explicit 
spatial evolution and before dealing with radiative and Coulomb coupling, the 
change in the nonthermal electron spectrum $n(\gamma_j)$ over a proper time 
interval $\Delta\tau$ at each bin $j$ in $\gamma$-space is 
 
 \begin{equation}
 \label{eq::adiabupwind}
 \Delta n(\gamma_j) = \Delta \tau 
\left(\frac{u^\alpha_{\;\;;\alpha}}{3}\right)\left[\frac{\d}{\d\gamma}
\left((\gamma-\gamma^{-1})n(\gamma)\right)\right]_j.
\end{equation}

 The expansion parameter $u^\alpha_{\;\;;\alpha}$ is computed from the 
$u^\alpha$ obtained at the end of the explicit operator. For numerical 
stability, the derivative $\d/\d\gamma$ is approximated using explicit upwind 
finite differencing. The upwind direction depends on the sign of the expansion.
 
 Because the upwind evolution in equation~\eqref{eq::adiabupwind} conserves 
total particle number but not energy, the spectrum $\Delta n(\gamma)$ of 
particles added or subtracted to the distribution is scaled so that the total 
change in  energy is equal to the amount predicted by equation~\eqref{eq::energynorm} 
(see Section~\ref{sec::energycons}).
 
 \item If the expansion $u^\alpha_{\;\;;\alpha} > 0$, nonthermal electrons may 
escape out of the lowest bin of the distribution. The loss of energy and 
particles out of the lowest bin is calculated and added to the thermal 
distribution number and energy density. Similarly, if $u^\alpha_{\;\;;\alpha}<0$, nonthermal electrons may escape out of 
the highest bin, and the corresponding flux of energy and number density is 
added to the thermal distribution. The thermal electron entropy per particle 
$s_e$ is recomputed using the updated number and energy density. 

 \item Since each species has now gone through its full adiabatic evolution, we 
calculate the viscous dissipation rate $q^v$ in each cell by comparing the total 
fluid energy density after the explicit step with the sum of the current species 
energies \citep{KORAL16},
 \begin{equation}
 \label{eq::vischeat}
  q^v = \frac{1}{\Delta\tau}\left(u - u_{i\,\text{th}\,\text{adiab}} - 
u_{e\,\text{th}\,\text{adiab}} - u_{e\,\text{nth}\,\text{adiab}}\right).
 \end{equation}
 Here, $u$ is the internal energy density of the total gas after the explicit 
step over a fluid frame proper time step $\Delta\tau$. 
$u_{i\,\text{th}\,\text{adiab}},u_{e\,\text{th}\,\text{adiab}}$,and 
$u_{e\,\text{nth}\,\text{adiab}}$ are the internal energy densities carried by thermal 
ions, electrons, and nonthermal electrons after adiabatic evolution. The 
difference between $u$ and the sum of the adiabatically evolved species energy 
densities gives the total energy gained from viscous dissipation during the time 
step.
 
 \item The fraction of the viscous heating applied to the electrons $\delta_e$, 
and the fraction of that applied to the nonthermal population 
$\delta_\text{nth}$ are calculated depending on the prescription used.

 \item Particles are added to the nonthermal population in a power-law 
distribution by adding the quantity in equation~\eqref{eq::powlawinj} to each 
bin. 
 
 \item The thermal energy densities $u_\text{e,\,th}$ and $u_i$ are
   increased by their fraction of the remaining viscous heating. The
   corresponding changes in thermal entropies $s_e$ and $s_i$ are computed
   by equations~(\ref{eq::adiabspecies}--\ref{eq::entpp}).
\end{enumerate}
 
\begin{figure*}
\centering
\includegraphics*[width=\textwidth]{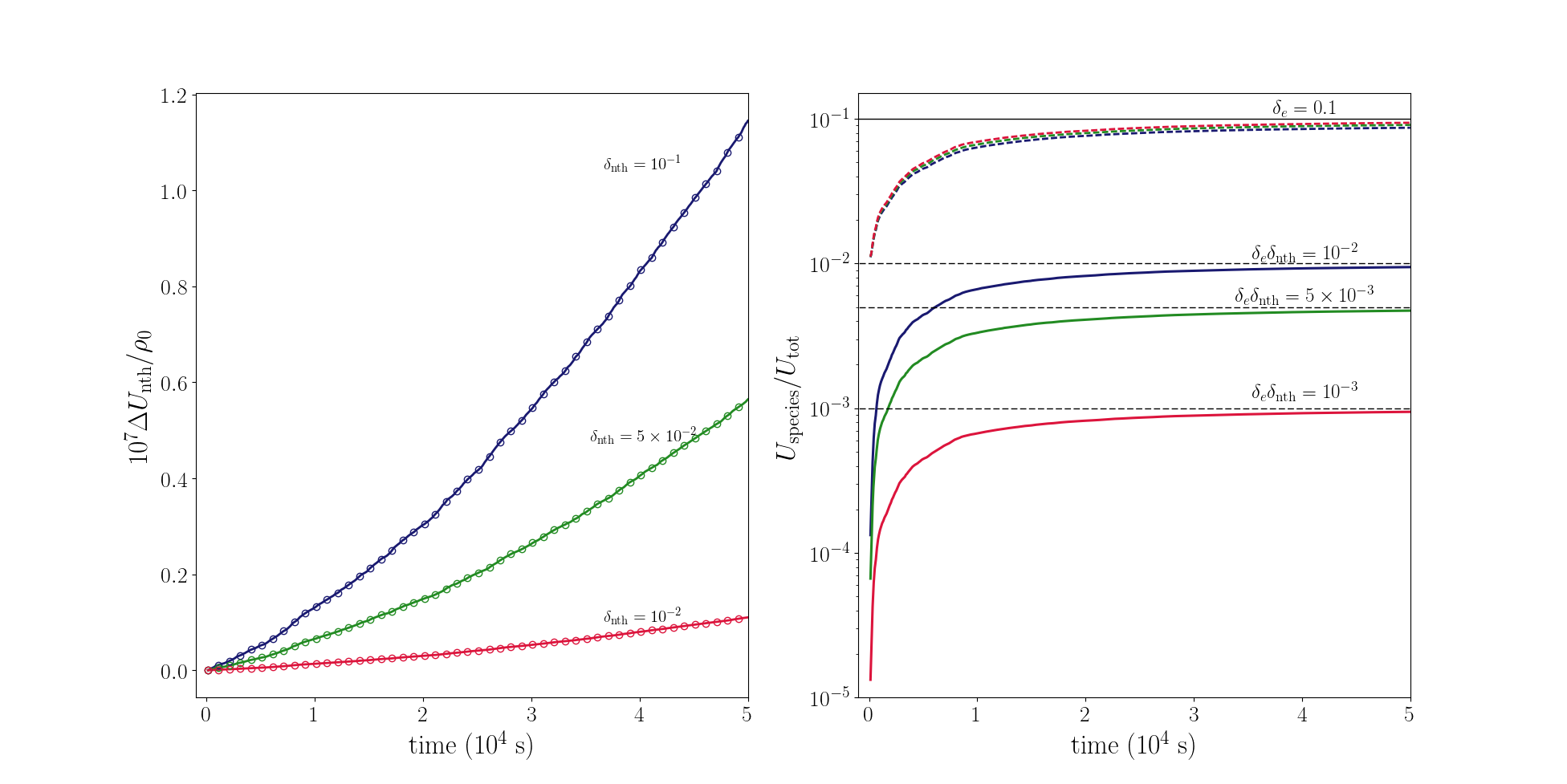}
\caption
{(Left) The increase of the total energy of nonthermal electrons
  $\Delta U_\text{nth}$ integrated over the turbulent box of
  Section~\ref{sec::turbtest}. The total electron heating fraction was
  set at $\delta_e=0.1$, and three runs were performed with nonthermal
  heating fraction $\delta_\text{nth}=0.01$ (red),
  $\delta_\text{nth}=0.05$ (green), and $\delta_\text{nth}=0.1$
  (blue). The open circles indicate the increase of the internal
  energy of the nonthermal population, and the solid lines show the
  predicted increase, which is the fraction
  ($\delta_e\delta_\text{nth}$) of the increase in the total gas
  energy.  (Right) The fraction $U_\text{species}/U_\text{gas}$ of
  thermal electrons (dashed lines) and nonthermal electrons (solid
  lines). As time proceeds and energy from viscous dissipation is
  divided among the different species, the energy fraction in each
  species asymptotes to the value given by the corresponding fixed
  viscous heating injection fractions:
  $\delta_e(1-\delta_\text{nth})$ for thermal electrons, and
  $\delta_{e}\delta_\text{nth}$ for nonthermal electrons.  }
\label{fig::turbtest}
\end{figure*}

\begin{figure*}
\centering
\includegraphics*[width=\linewidth]{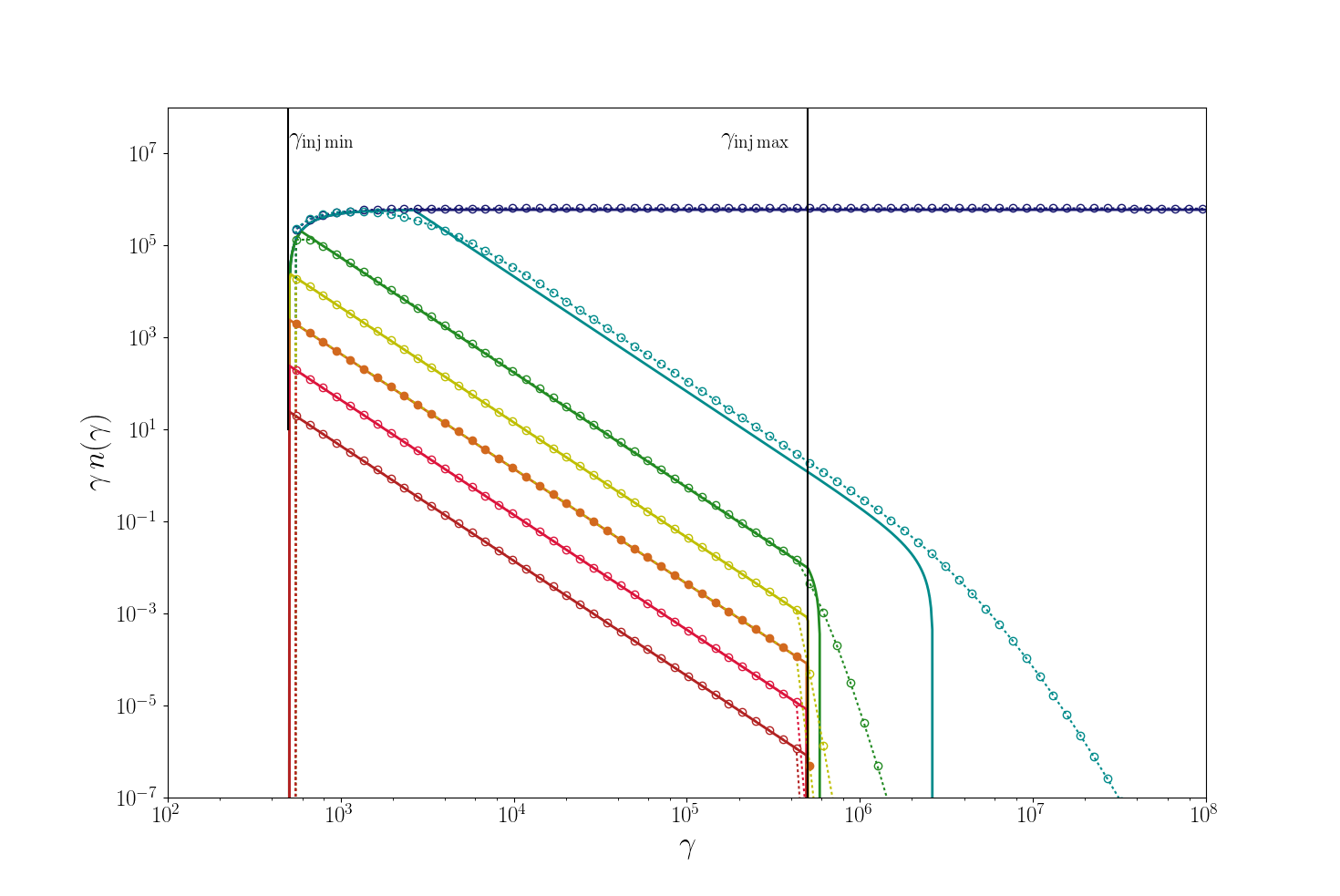}
\caption
{ Results of a test of adiabatic compression with constant
  $u^\mu_{\;\; ; \mu}=-5\times10^{-3}$ s$^{-1}$ and particle injection
  with slope $p=3.5$ between $\gamma_{\rm{inj \, min}}=50$ and
  $\gamma_{\rm{inj \, max}}=5\times10^5$. The injection distribution
  is normalized so that the total injection rate is 1000 particles
  cm$^{-3}$ s$^{-1}$. The solid lines show the analytic solution to
  the problem at times (from below) $t=10^{-2}, 10^{-1}, 1, 10, 10^2,
  10^3$ and $10^4$ seconds. The open circles show the \texttt{KORAL}
  solution at the corresponding times.
}
\label{fig::adiabheattest}
\end{figure*}

\subsection{Implicit Radiation and Coulomb Coupling}
\label{sec::implicitev}

The source terms representing the radiative and Coulomb coupling
between the species are: the radiative coupling $G_\nu$, the thermal
Coulomb coupling $q^C_\text{th}$, the photon source term $\dot{n}_r$,
the nonthermal cooling rates $\dot{\gamma}$, the nonthermal Coulomb
coupling $q^C_\text{nth}$, and the cooling from the nonthermal population to the
thermal bath, $q^{\text{cool}}$ and $\mu\dot{n}^\text{cool}$. These
coupling terms in equations~\eqref{eq::GRRMHD}, \eqref{eq::ent_ev} and
\eqref{eq::elecev} are applied through a semi-implicit operator using
the methods described in \citet{KORAL13, KORAL14}.
 
The implicit solver uses a reduced set of primitives which includes the energy 
density and velocity of \emph{either} gas or radiation, the photon number 
density, electron energy density, and the full nonthermal distribution. The 
other primitives, including the velocity not evolved, the gas density, and the 
ion entropy are continually updated during the iterations of the implicit solver 
to enforce the appropriate conservation laws.
  
\subsection{Energy vs Particle Conservation}
\label{sec::energycons}

The equation that governs the evolution of the nonthermal
distribution, equation~\eqref{eq::elecev}, is a conservation law for a
particle current in five dimensions, three spatial, one time, and one
corresponding to the individual particle energies. As this equation is
evolved via our finite volume algorithms (and as we account properly
for the loss of particles at $\gamma_{\rm{min}}$ and
$\gamma_{\rm{max}}$), the total number of particles in the distribution
is conserved. The total internal energy density in the distribution is
given by integrating $n(\gamma)$ times the particle energy
$(\gamma-1)m_e c^2$ over $\gamma$ (equation~\ref{eq::uthint}). While
particles are not lost from the distribution (excepting boundary
effects), energy can be lost in radiative cooling, Coulomb coupling,
or adiabatic expansion; it can be gained in adiabatic
compression. Because the finite volume form of
equation~\eqref{eq::elecev} does not conserve the particle energy
current, and because we use a numerical approximation to the integral
in equation~\eqref{eq::uthint} to compute the internal energy in the
nonthermal distribution, the evolution of the nonthermal distribution
on its own does not conserve energy.

We account for this in two ways. In the implicit step, where nonthermal 
electrons lose energy to radiation and Coulomb coupling, we simply ensure that 
overall energy is conserved by adjusting the nonthermal energy flux into 
radiation ($-\hat{G}^0_\text{nth}$) to reflect the energy that is actually lost 
in cooling the nonthermal distribution. That is, instead of using 
equation~\eqref{eq::g0nth}, we compute $\hat{G}^0_\text{nth}$ by computing the 
difference in the total nonthermal energy density at a given substep in the 
implicit solver with the energy density computed before the implicit step, 
subtracting off the small part of the cooling that is due to Coulomb coupling 
(which goes into thermal electrons). When we need to compute 
$\hat{G}^0_\text{nth}$ outside of the implicit solver, we use 
equation~\eqref{eq::g0nth}. In this way, the total energy is conserved and the 
shape of the nonthermal distribution is not affected, although the total energy 
in the distribution may differ from the value computed from an analytic solution 
or found in a simulation with finer sampling in $\gamma$.

In the intermediate step, where particles are heated or cooled by
adiabatic compression or expansion, we cannot account for the
missing/extra energy by simply adding it to
radiation or the thermal population. This is because the adiabatic
heating/cooling of the nonthermal distribution is part of the
adiabatic evolution, and correctly computing the viscous heating via
equation~\eqref{eq::vischeat} depends on properly evolving the
independent species energies adiabatically. In this case, we evolve
the distribution explicitly, and then scale the computed $\Delta
n(\gamma_j)$ at each sampled $\gamma_j$ so that the total change in
energy is equal to the amount given by the total instantaneous rate of
energy increase,
\begin{equation}
\label{eq::energynorm}
 \Delta u = m_e c^2 \int \dot{\gamma}_\text{adiab} n(\gamma) \rm{d}\gamma.
\end{equation}
Scaling the distribution in this way can bias the shape of the distribution 
(Section~\ref{sec::adiabtest}). However, the energy gained and lost in this 
step is applied correctly (Section~\ref{sec::turbtest}), and the computation of 
the viscous heating rate $q^v$ is consequently not biased.

\section{Tests}
\label{sec::test}

In this section we describe several test problems to demonstrate the accuracy of 
the nonthermal electron evolution as implemented in \texttt{KORAL}. In Section 
\ref{sec::turbtest}, we test the bulk heating of the nonthermal electrons in a 
turbulent box. In Section~\ref{sec::adiabtest} we compute the evolution of the 
electron distribution under constant injection and adiabatic compression, and in 
Section~\ref{sec::syntest} we test evolution from injection, synchrotron 
cooling, and inverse Compton scattering. We compare the numerical results with 
analytic and semi-analytic solutions to gauge the accuracy of the algorithms. 

\begin{figure*}
\centering
\includegraphics*[width=\linewidth]{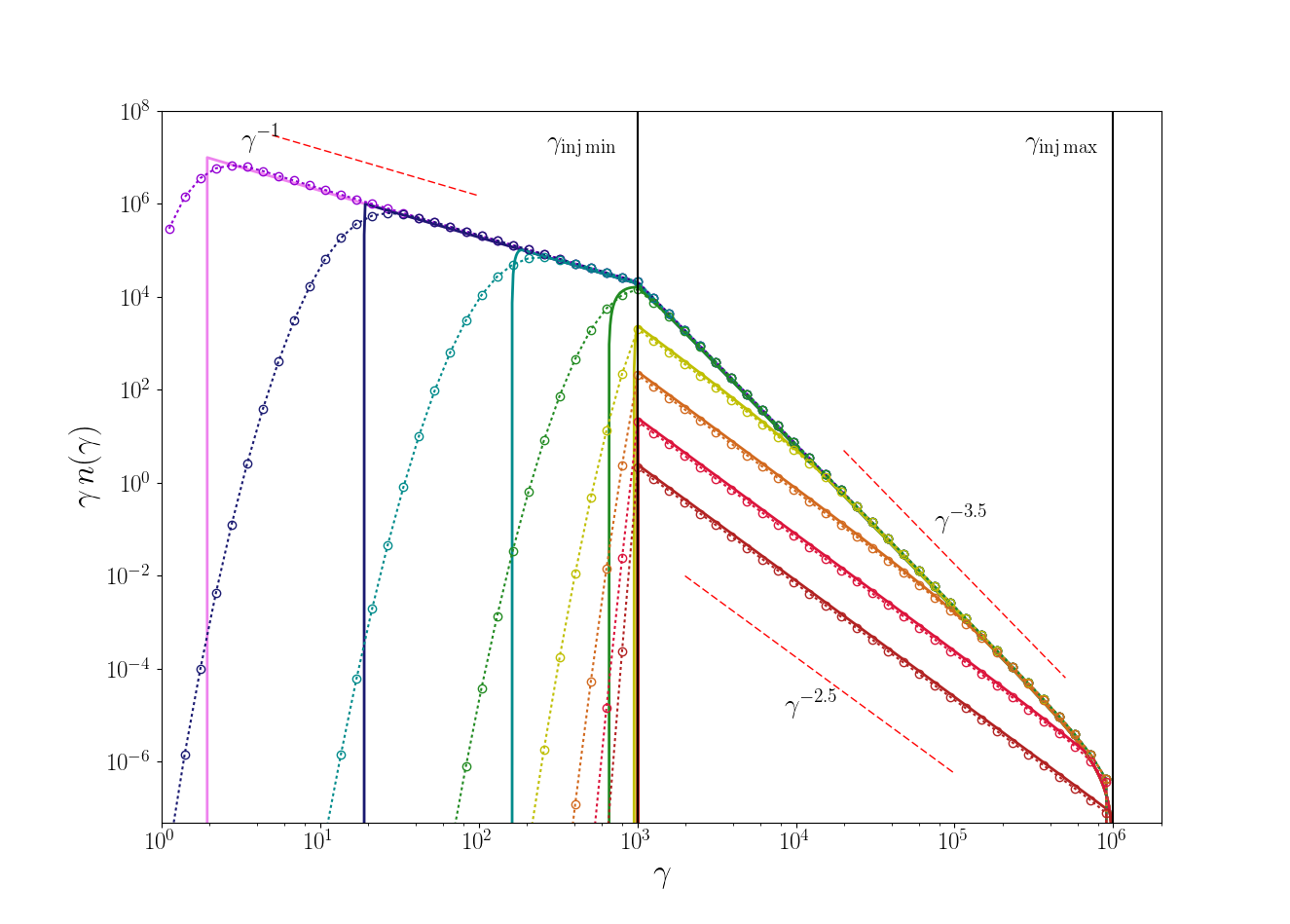}
\caption
{Results of a test with constant particle injection with slope
  $p=3.5$ between $\gamma_{\rm{inj,min}}=50$ and $\gamma_{\rm{inj,max}}=5\times10^5$ 
  coupled with synchrotron cooling in a uniform
  magnetic field with $B=200$ G. The injection distribution is
  normalized so that the total injection rate is 1000 particles
  cm$^{-3}$ s$^{-1}$. The numerical solution from \texttt{KORAL}'s
  implicit solver (open circles) is compared with the analytic
  solution (solid lines) at times $t=10^{-3}, 10^{-2}, 10^{-1}, 1, 10,
  10^2, 10^3$ and $10^4$ seconds. The spectrum develops a cooling
  break between the injection slope $p$ for $\gamma <
  \gamma_\text{brk}$ and $p+1$ for $\gamma>\gamma_\text{brk}$. The
  cooling break starts at large $\gamma$ and propagates toward lower
  $\gamma$ until the spectrum is broken over the entire injection
  range at $t=10$ s. After this time, the spectrum cools to 
  $\gamma<\gamma_{\rm{inj,min}}$ with slope $p=2$. The sharp
  discontinuity at the lower end of $n(\gamma)$ is smeared out in the
  numerical \texttt{KORAL} solution because of diffusion in the upwind
  finite differencing method we use. However, \texttt{KORAL}
  accurately captures the location of the peak of $\gamma n(\gamma)$
  as it propagates to lower energies.  }
\label{fig::syncooltest}
\end{figure*}

\begin{figure*}
\centering
\includegraphics*[width=\linewidth]{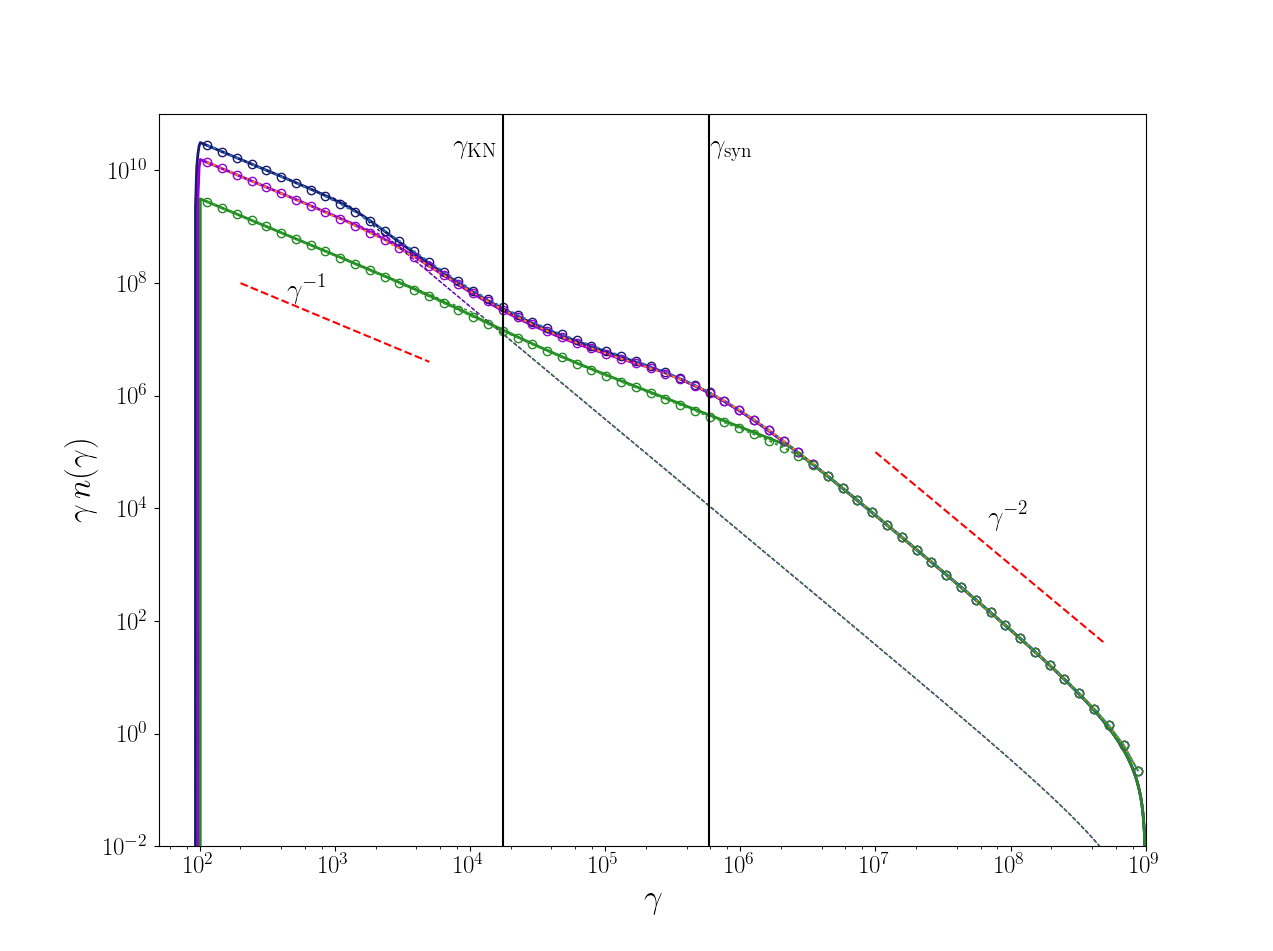}
\caption
{Nonthermal energy distribution evolution in an environment with $B=10
  \,\mu$G, $u_\text{rad}=8.01\times10^{-10}$ erg cm$^{-3}$, and
  $T_\text{rad}=30000$ K. We inject particles in a power law with
  $p=2$ between $\gamma_{\rm{inj,min}}=100$ and $\gamma_{\rm{inj, max}}=10^9$. 
  The numerical solution from \texttt{KORAL}'s
  implicit solver (open circles) is compared with the semi-analytic
  solution (solid lines) at times $t=10^5$ (green), $5\times10^5$
  (purple) and $10^6$ yr (blue). Note the excellent agreement. The
  analytic solution for the same problem neglecting the Klein-Nishina
  cross section of electrons (taking $F_{KN}=1$ in
  equation~\ref{eq::iccoolrate}) is also displayed (dotted lines).
}
\label{fig::iccooltest}
\end{figure*}

\subsection{Driven Turbulence}
\label{sec::turbtest}

To test the implementation of adiabatic evolution and viscous heating of the 
nonthermal population, we repeated the turbulent box test from \citet{KORAL16}, 
which was inspired by the MHD driven turbulence test of \citet{Ressler15}. A 
fraction $\delta_e = 0.1$ of the dissipative heating $q^v$ is deposited into the 
electrons, of which a fraction $\delta_\text{nth}$ goes into the nonthermal 
population via equation~\eqref{eq::powlawinj}, with the remaining fraction 
$\delta_e(1-\delta_\text{nth})$ going into the thermal electrons by 
equation~\eqref{eq::ent_ev}. 

We start with an initial uniform two dimensional system of size $L$ with density 
$\rho_0$, zero velocity, speed of sound $c_{s0}=8.6\times10^{-4} c$, a 
horizontal magnetic field with $\beta=p_\text{gas}/p_\text{mag}=6$, no 
nonthermal electrons, and periodic boundary conditions. We drive the system with 
random, divergence-free Gaussian perturbations in the velocity with a power 
spectrum $P(|\delta v|^2) = k^6\exp(-8k/k_\text{pk})$, where $k_\text{pk} = 
4\pi/L$. These perturbations add kinetic energy to the system which dissipates 
into internal energy of the gas, divided among the three species. Radiation and 
Coulomb coupling are turned off.

The open circles in the left panel of Fig.~\ref{fig::turbtest} shows
the resulting increase of the total energy in nonthermal electrons
integrated over the simulation volume for three runs with
$\delta_\text{nth}=.01,.05$,and $.1$, respectively (open circles). The
circles are compared with the corresponding fraction
$\delta_e\delta_\text{nth}$ of the increase in the total gas energy
(solid lines). The close agreement shows that the combination of
viscous heating and the net change in energy from adiabatic
compression and expansion (as a result of turbulence) is handled
correctly. In particular, the energy normalization performed on the
nonthermal distribution during the adiabatic compression/expansion
step (Section ~\ref{sec::energycons}) is necessary to identify the
correct amount of viscous heating and produce the good agreement shown
in Fig.~\ref{fig::turbtest}.

The right panel of Fig.~\ref{fig::turbtest} shows the ratio of the
energy densities of the two electron populations to the total gas
energy density: $U_\text{th}/U_\text{gas}$, and
$U_\text{nth}/U_\text{gas}$. As energy is dissipated and divided among
the species, the ratios of the species energies to the total internal
energy correctly asymptote to the injection fractions
$\delta_e(1-\delta_\text{nth})$ and $\delta_{e}\delta_\text{nth}$ for
thermal and nonthermal electrons, respectively.

\subsection{Particle Injection and Adiabatic Compression}
\label{sec::adiabtest}

To test the implementation of the adiabatic heating and cooling of
electrons under gas compression and expansion
(equation~\ref{eq::gdotadiab}), we consider a zero-velocity gas
background with constant injection of nonthermal electrons with a
power-law slope $p=3.5$ between $\gamma_{\rm{inj \, min}}=50$ and
$\gamma_{\rm{inj \, max}}=5\times10^5$. We also subject the system to
a constant artificial compression rate (not computed from the actual gas four-velocity)
$u^\mu_{\;\; ; \mu}=-5\times10^{-3}$ s$^{-1}$, 
similar to the compression rate found in the equatorial
plane at a radius of $~\sim 5\, r_g$ in the accretion disc simulations
described later in Section~\ref{sec::sgra}. We turn off the radiative
and Coulomb coupling interactions. The analytic solution to this
problem (\citealt{Mano}, Appendix A) shows the development of a break
from the injection power-law slope $-p$ to a slope of $-1$ at low
$\gamma$, with the break propagating to higher $\gamma$ with
increasing time.

Fig.~\ref{fig::adiabheattest} shows the results of the test at
logarithmically spaced time intervals. The open circles, which denote
the \texttt{KORAL} solution, mostly line up well with the analytic
result. Deviations arise from two effects. First, the numerical scheme
is diffusive and thus smooths out sudden breaks in the slope of
$n(\gamma)$. This is seen as a tail above the maximum $\gamma$ of the
true distribution, and also at the break between slope -3.5 to -1,
around $\gamma=3000$ for $t=1000$ s. Second, the smoothing out of breaks leads to 
a loss of energy from the analytic value. Since we conserve the total energy, this leads to a shift in
the normalization (Section~\ref{sec::energycons}). This effect is
obvious in the \texttt{KORAL} solution at $t=10^3$ s. Once the
spectrum has broken completely, the \texttt{KORAL} solution matches
the analytic solution for all $\gamma$. In practice, fluid in a
turbulent simulation will experience many phases of compression and
expansion, which should wash out the energy correction effect
illustrated in this test.

\subsection{Synchrotron and Inverse Compton Cooling}
\label{sec::syntest}

We check the implementation of radiative cooling in \texttt{KORAL}'s implicit 
solver with two tests in a flat, zero-velocity gas background with constant 
injection of nonthermal electrons. 

In the first test, we inject particles with a power-law slope $p=3.5$ between 
$\gamma_{\rm{inj \, min}}=50$ and $\gamma_{\rm{inj \, max}}=5\times10^5$ and 
subject them to synchrotron cooling in a constant magnetic field of $B=200$ G.  
Under constant injection and synchrotron cooling, and for $t<t_\text{syn}$, the 
particle spectrum develops a cooling break from the injection power-law slope 
$-p$ to a slope $-(p+1)$ at $\gamma_\text{brk}$ given by
\begin{align}
 \label{eq::synbrk}
 \gamma_\text{brk}&=(1/\gamma_{\rm{inj \, max}} - b_s t)^{-1}, \;\;\; \\ 
\nonumber
 b_s &= 1.292\times10^{-9}(B/\,1\,\text{G})^2,
\end{align}
where the time $t_\text{syn}$ is
\begin{equation}
\label{eq::syncooltime}
t_\text{syn} = (\gamma_{\rm{inj \, min}}^{-1}-\gamma_{\rm{inj \,  
max}}^{-1})/b_s.
\end{equation}
At $t=t_\text{syn}$, the cooling break reaches $\gamma_{\rm{inj \, min}}$, and 
at later times the spectrum cools to $\gamma < \gamma_{\rm{inj \, min}}$ with a 
power-law slope of $-2$. 

The results from \texttt{KORAL} are compared with the analytic solution in 
Fig.~\ref{fig::syncooltest}. The development of the synchrotron cooling break 
and its propagation to lower particle energies with time is clearly captured in 
the \texttt{KORAL} solution. At the low resolutions we use, the numerical 
solution from \texttt{KORAL} cannot capture the sharp cutoff at low particle 
energies, and produces a tail extending to low $\gamma$ (note that the vertical 
scale is over 14 orders of magnitude, so the discrepancy is not serious). 
However, the location of the peak in the spectrum as a function of time is 
reproduced well.

As another test of the \texttt{KORAL} implicit solver for radiative nonthermal 
cooling, we replicate a problem from \citet{Mano}, which demonstrates the 
effects of the Klein-Nishina cross section in the inverse Compton cooling term 
(equations.~\ref{eq::iccoolrate} and ~\ref{eq::FKN}). Neglecting bremsstrahlung 
radiation and Coulomb coupling, the cooling rate is
\begin{equation}
 \label{eq::ictestgdot}
 \dot{\gamma} = b_\text{syn}\gamma^2\left[1 + 
\frac{u_\text{rad}}{u_\text{mag}}\left(1+4\gamma\epsilon_0\right)^{-3/2}\right],
\end{equation}
where $b_\text{syn} = -1.292\times10^{-11}\left(B/1\,\text{G}\right)^2$ and 
$\epsilon_0 = kT_r/m_ec^2$. We set up a test in a uniform background 
similar to a stellar environment dominated by hot young stars \citep{Mano}. We 
set $B=10\; \mu$G and assume a photon bath with energy density 
$u_\text{rad}=7.95\times10^{-10}$ erg cm$^{-3}$  and temperature 
$T_\text{rad}=30000$ K. We inject particles in a power law with $p=2$ between 
$\gamma_{\rm{inj \, min}}=100$ and $\gamma_{\rm{inj \, max}}=10^9$, normalized 
so that the total injection rate is 10$^{-3}$ particles cm$^{-3}$ s$^{-1}$.  

We compute the spectrum as a function of time using the semi-analytic method of 
\citet{Mano}. The results are displayed in Fig.~\ref{fig::iccooltest} at times 
$t=10^5, \, 5\times10^5$, and $10^6$ yr. The \texttt{KORAL} solution (open 
circles) lines up well with the semi-analytic solution (solid lines), 
demonstrating the code's ability to accurately capture details of the radiative 
cooling of nonthermal distributions beyond simple synchrotron cooling. 

The solution in this test displays different behavior in three distinct regimes. 
From equation~\eqref{eq::ictestgdot}, at the highest energies, 
$\gamma>\gamma_\text{syn} = \left((u_\text{rad}/u_\text{mag})^{2/3} - 
1\right)/4\epsilon_0$, the solution is dominated by synchrotron cooling. Hence 
the spectrum shows a characteristic synchrotron cooling break above 
$\gamma_\text{syn}$, where the slope  becomes $-(p+1)=-3$. 
Equation~\eqref{eq::ictestgdot} also indicates that below $\gamma_{KN} \approx 
1/4\epsilon_0$, the Thomson limit applies. Between $\gamma_{KN}$ and 
$\gamma_\text{syn}$, the decrease in the cooling rate due to the Klein-Nishina 
cross section causes the spectrum to harden compared to what is predicted when 
only Thomson scattering is considered (dotted lines in 
Fig.~\ref{fig::iccooltest}). As time progresses, electrons initially injected at 
$\gamma_{\rm{inj,max}}$ cool to lower energies $\gamma_\text{cool}$. By the last 
time shown, $\gamma_\text{cool} < \gamma_{KN}$; electrons injected at the 
highest energies have cooled below the energies where the Klein-Nishina cross 
section dominates, and Thomson cooling begins to break the spectrum for $\gamma 
< \gamma_{KN}$. 

Because the cooling rates in this problem are so low, even over $10^6$
years the spectrum does not have time to cool much below the injection
range. Therefore, the implicit solver does not have to deal with
abrupt discontinuities in the spectrum, and except for the slight
smoothing out of the synchrotron break, the obvious
diffusion seen in the tests in Figs.~\ref{fig::adiabheattest} and
\ref{fig::syncooltest} is not apparent here.

\section{Test Simulation of Sgr A*}
\label{sec::sgra}

\begin{figure*}
\includegraphics*[width=\linewidth]{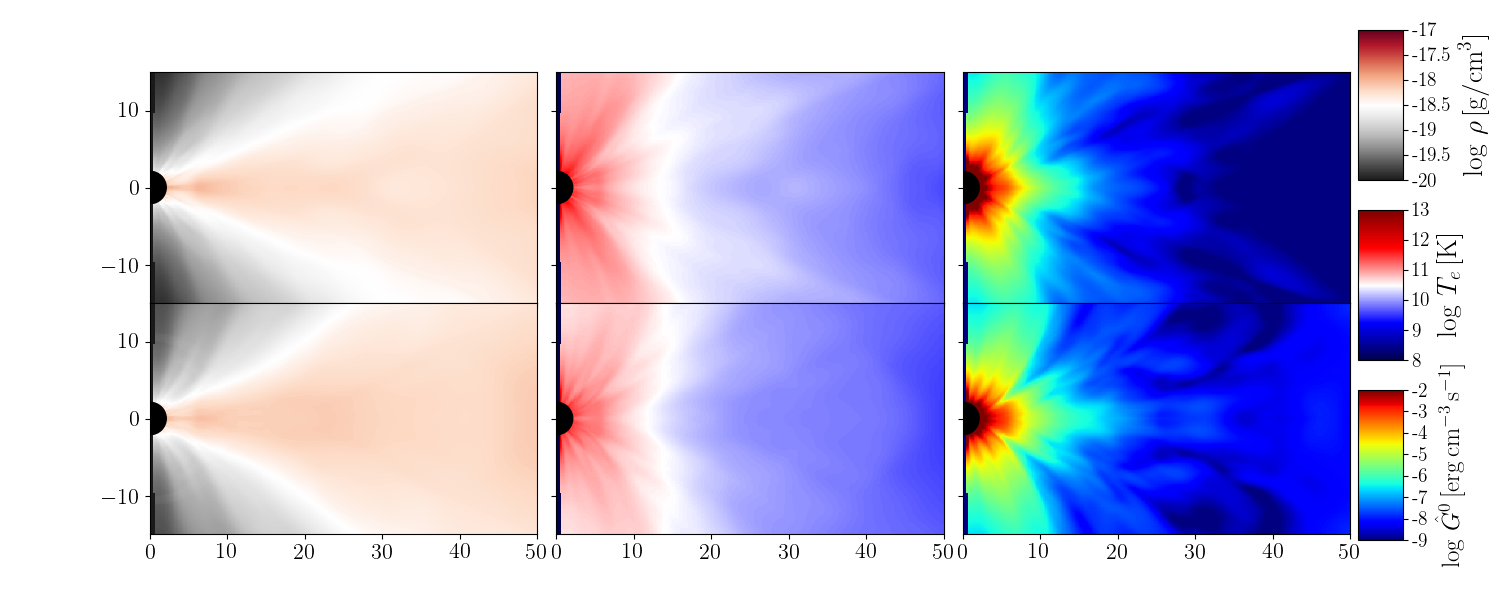} 
\caption{Comparison of time-averaged quantities in the control thermal
  run (top row) and the nonthermal run (bottom row).  Averages are
  taken over the period $t=1.5 \times 10^4 -2\times10^4\,t_g$, and the
  distributions are symmetrized about the equatorial plane. In each
  column, the same color scale is used in the upper and lower
  panels. From left to right, the quantities shown are the gas density
  $\rho$, the electron temperature $T_e$, and the fluid frame
  radiation power $-\hat{G}^0$. The
  presence of nonthermal electrons does not significantly affect
  either $\rho$ or $T_e$. However, the nonthermal model has
  significantly more radiative power, especially at larger radii.}
\label{fig::compplots}
\end{figure*}

As a test of the entire code with all elements included, we have
carried out 2D simulations of an accreting black hole with parameters
appropriate for the accretion flow in \sgraa.  We have run both a pure
thermal model with a two-temperature plasma (two fluid populations,
thermal ions and thermal electrons, similar to \citealt{KORAL16}), and
a nonthermal model with all three fluid populations (thermal ions,
thermal electrons and nonthermal electrons, \S\ref{sec::fluidpops}).

For the injection properties of the nonthermal population we use a
very simple ad hoc prescription, which is sufficient for the current
test, but will need to be improved in the future for modeling real
systems.  We assume that a constant fraction of the local viscous
electron heating rate goes into nonthermal electrons, with a fixed
energy spectrum that is independent of location in the simulation
box. However, the observed infrared and X-ray variability of \sgra
\citep{Dodds11, Neilsen} suggests that the nonthermal acceleration
mechanism is localized, either in magnetic reconnection regions
\citep{Sironi14} or in shocks \citep{Guo14}. Recently, \citet{Ball_16} showed 
that the X-ray variability of \sgra could be qualitatively
reproduced by adding a nonthermal distribution by hand in regions of
high magnetization in a single-fluid GRMHD simulation. In a future
work, we will consider full GRMHD+nonthermal electron simulations
using more elaborate injection prescriptions informed by these studies.

\subsection{Units}
\label{sec::units}

We work in a Schwarzschild spacetime (non-spinning black hole) with
black hole mass $M=4\times10^6 \, M_\odot$, the estimated mass of
\sgra \citep{Gillessen09,Chatzopoulos15}. We use the gravitational
radius $r_g = GM/c^2 = 6\times10^{11} \,\text{cm} = 0.04 \,\text{AU}$
as our unit of length, and $t_g = r_g/c = 20$\,s as our unit of time. 
We define the Eddington accretion rate as
\begin{equation}
 \dot{M}_\text{Edd} = \frac{L_\text{Edd}}{\eta c^2} = \frac{4\pi GMm_p}{\eta c 
\sigma_T},
\end{equation}
where $L_\text{Edd}$ is the Eddington luminosity, and we use an
efficiency $\eta = 0.057$ appropriate for a non-spinning black
hole. For the $4\times10^6 M_\odot$ black hole we consider, the
Eddington accretion rate $\dot{M}_\text{Edd}=0.16 \, M_\odot$
yr$^{-1}$, and the Eddington luminosity
$L_\text{Edd}~=~5\times10^{44}$ erg s$^{-1}$.

\subsection{Model Setup}
\label{sec::setup}

The simulations are performed in Kerr-Schild coordinates using an
axisymmetric 2D grid with a resolution of $256\times256$ cells in
radius and polar angle. The radial cells are distributed exponentially
from inside the BH horizon at $1.85 \, r_g$ to 1000 $r_g$, and the
polar angle cells are uniformly sampled.

The initial fluid conditions are identical to the model
\texttt{Rad8SMBH} in \citet{KORAL16}. We initialize the simulation
with a hydrostatic equilibrium torus with parameters as in
\citet{Narayan2012}. The torus has an inner edge at $10 \, r_g$ and is
threaded by a weak magnetic field. The initial electron and ion
temperatures are set equal to the initial gas temperature, and there
are no nonthermal electrons. The torus is surrounded by a
static atmosphere with negligible mass and radiation energy
density, but with the radiation temperature everywhere set to $10^5$
K.

We ran the model for a total time of $2\times 10^4 \, t_g$ with
nonthermal electron evolution turned off. The thermal electron and ion
populations were heated using the viscous heating prescription of
\citet{Howes10}. They exchanged energy with each other via thermal Coulomb
coupling, and the thermal electrons radiated via synchrotron,
bremsstrahlung, and inverse Compton scattering. We used the mean-field
dynamo from \citet{KORAL15} to prevent the decay of the axisymmetric
magnetic field.  We refer to this purely thermal simulation as the
{\it control run}.

In the control run, gas begins accreting on the black hole around
$t\sim3000\,t_g$, and by $t=10^4\,t_g$, the accretion is in
steady state and we can estimate the mass accretion rate on the black
hole.  At this time, we scale the gas density, and correspondingly the
magnetic field strength, to achieve the desired accretion rate of
$\approx 4\times10^{-8} \dot{M}_\text{Edd}$ appropriate for \sgraa. We
run the model with the rescaled density from $t=10^4\,t_g$ up to
$2\times10^4\,t_g$.  We use the data from the time period
$1.5\times10^4-2\times10^4\,t_g$ to study the properties of the
accretion flow.

Having run the control model described above, we then simulate a
system with nonthermal electrons included.  We refer to this
simulation as the {\it nonthermal run}.  We start this simulation with
the output from the control run at time $10^4\,t_g$, and rescale the
density and magnetic field as before.  However, we now include
nonthermal electron injection, and we evolve the system from $t=10^4\,t_g$
up to $2\times10^4\,t_g$ with all the nonthermal interactions turned
on. We track the nonthermal electron energy distribution over $N=32$ bins
ranging from $\gamma_\text{min}=200$ to
$\gamma_\text{max}=2\times10^6$. We chose $\gamma_\text{min}$ to be above
the characteristic electron energy $\theta = k_B T/m_ec^2$ for a temperature
at the high end of the range observed in the control model, around $T\sim10^{12}$ K. 
We then chose $\gamma_\text{max}$ so as to cover four decades of $\gamma$ in the nonthermal distribution. 
With a resolution of 8 points per decade, this corresponds to a total of
$N=32$ nonthermal electron bins.

For the nonthermal injection, we fix the power-law index at $p=3.5$,
consistent with past studies \citep{Ozel2000, Yuan2003} and with
observational constraints \citep{Porquet08,Barr14}. We inject the
electrons between $\gamma_{\rm{inj\,min}}=500$ and
$\gamma_{\rm{inj\,max}}=\gamma_{\rm{max}}=2\times10^6$. We offset the minimum injected
$\gamma$ from the lowest Lorentz factor $\gamma_\text{min}$ tracked by
the code in order to prevent the immediate cooling of electrons
injected at $\gamma_{\rm{min}}$ back into the thermal population
(equation~\ref{eq::edgeesc}). We set the nonthermal heating fraction to
$\delta_\text{nth} = 0.015$ \citep{Ozel2000, Yuan2003, Ball_16, Mao2016}. The total 
electron heating fraction $\delta_e$, of which
$98.5$ per cent goes to the thermal species, is determined using the
prescription of \citet{Howes10}, which is a (strong) function of the
magnetization parameter $\beta = p_{\rm gas}/p_{\rm mag}$.

\subsection{Comparison of Thermal and Nonthermal Models}
\label{sec::comparison}

In Fig.~\ref{fig::compplots}, we compare time-averaged spatial
distributions of several quantities in the control (thermal) run with
those in the nonthermal run. For each model, we average the quantities
over the time range $t=1.5\times10^4-2\times10^4\,t_g$, and
also symmetrize around the equatorial plane for additional smoothing
of the results. Shown are the density $\rho$, electron temperature
$T_e$, and the fluid frame radiation power $-\hat{G}^0$ computed from
average primitives.

Fig.~\ref{fig::compplots} indicates that the overall structure and
distribution of the gas density and electron temperature are similar
in the two models. This is expected, since the fraction of electron
energy that goes into the nonthermal electrons is only
$1.5$ per cent. Furthermore, the accretion flow in our model is optically
thin and radiatively inefficient, so the emission from the nonthermal
electrons does not significantly alter the gas dynamics. Indeed, the
gas dynamics and electron and ion thermodynamics in both the control
run and the nonthermal run are quite similar to model
\texttt{Rad8SMBH} in \citet{KORAL16}.

The last column of Fig.~\ref{fig::compplots}, however, shows that the
rest frame power of the emitted radiation is not the same in the
control and nonthermal runs --- it is enhanced in the latter run, most
significantly at large radii. The spatial distribution of the nonthermal emission
is purely the result of the particular injection prescription we use. 
We inject nonthermal electrons with the same energy fraction $\delta_{\rm nth}=0.015$,
in the same power-law distribution, everywhere in the
simulation. In addition, the magnetic field strength is fairly
constant ($B\sim10$\,G) over most of the region of
interest. Therefore, the amount of nonthermal synchrotron emission is
directly proportional to the viscous heating rate of the gas.  On the
other hand, the thermal electron temperature varies substantially with
radius, falling to below $\sim10^{10}$K by a radius of $30 \, r_g$. Since
thermal synchrotron power varies as $T_e^2$, the thermal emission
falls rapidly with increasing radius. Thus, the thermal electrons are
more advection-dominated at large radii compared to the nonthermal
electrons.

\subsection{Nonthermal Simulation}
\label{sec::nonthermal}

\begin{figure*}
\centering
\includegraphics*[width=0.9\textwidth]{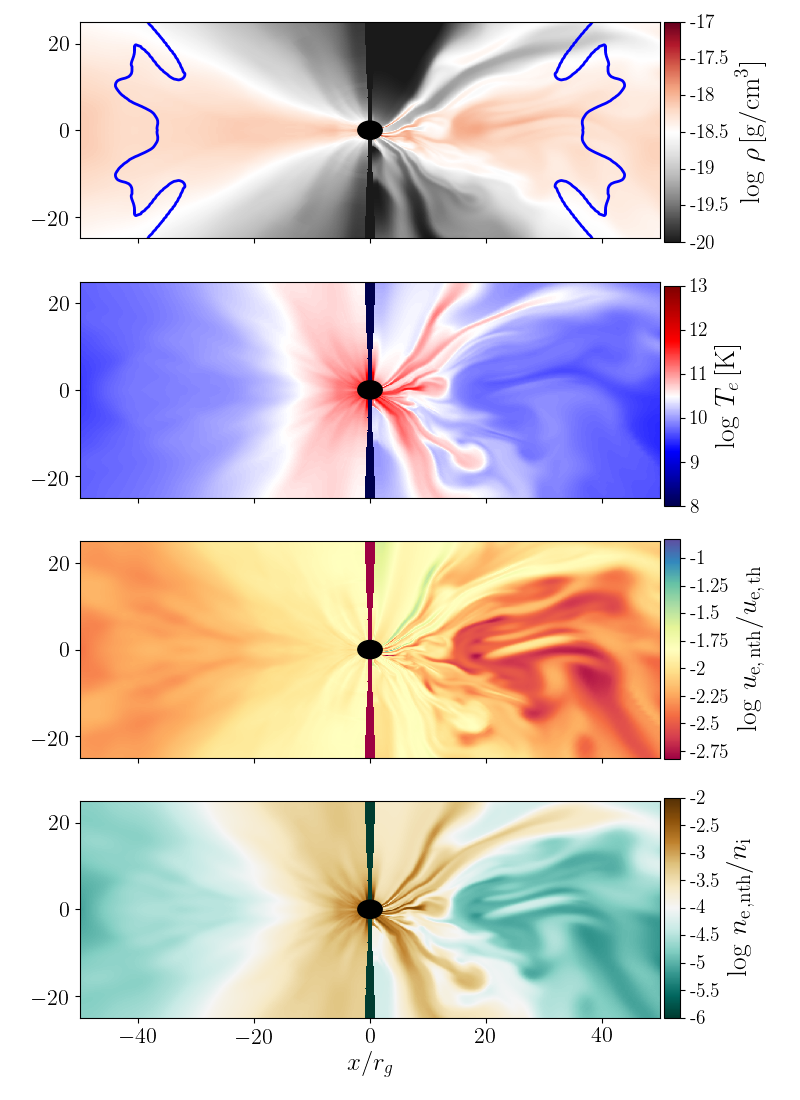}
\caption{Snapshot (right) and time-averaged (left) distributions of
  (from top to bottom) gas density $\rho$, thermal electron
  temperature $T_e$, ratio of nonthermal to thermal electron energy
  densities $u_\text{e,\;nth}/u_\text{e,\;th}$, and fraction of
  electrons in the nonthermal distribution
  $n_\text{e\;,th}/n_\text{i}$. The blue contour in the first panel
  encloses the region of the simulation that is in inflow equilibrium,
  as determined by equation \eqref{eq::tacc}.}
\label{fig::modelparams}
\end{figure*}

\begin{figure*}
\centering
\includegraphics*[width=0.9\textwidth]{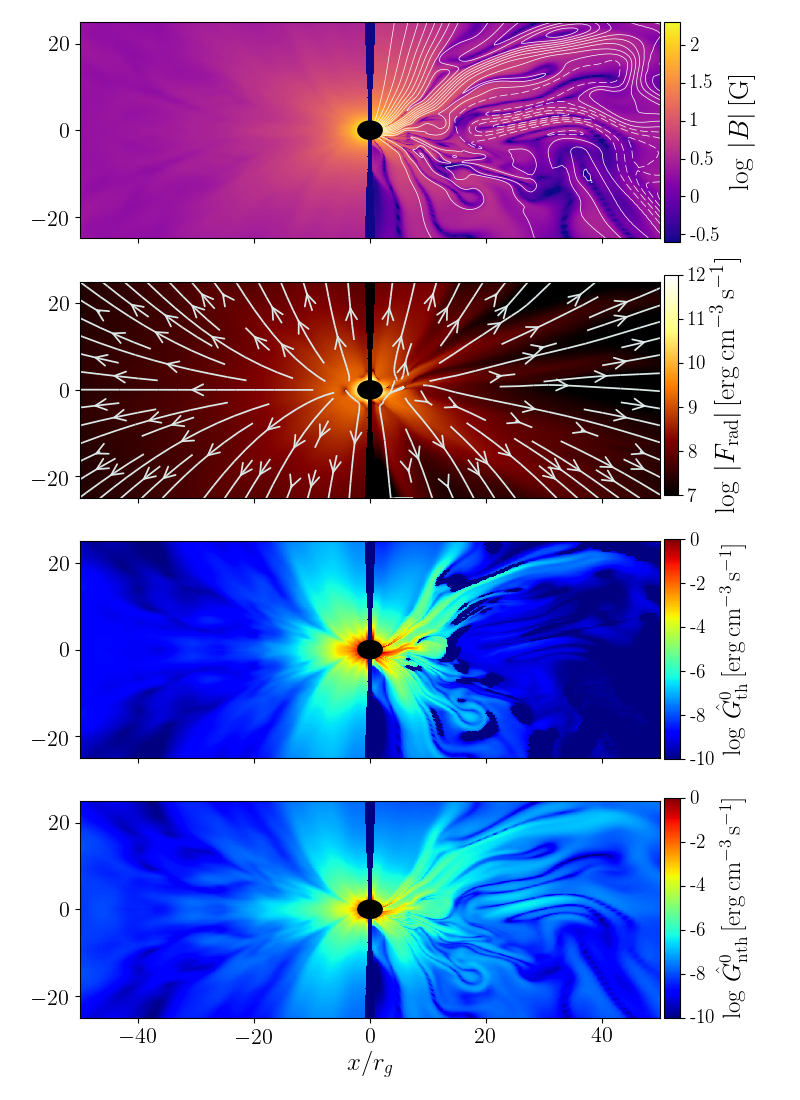}
\caption{Snapshot (right) and time-averaged (left) distributions of
  (from top to bottom) magnetic field strength $|B|$, magnitude of the
  radiation flux $|F|$, fluid frame radiation power from thermal
  electrons $-\hat{G}^0_\text{th}$, and fluid frame radiation power
  from nonthermal electrons $-\hat{G}^0_\text{nth}$.  Contours in the
  first panel show poloidal magnetic field lines. Streamlines in the
  second panel show the direction of the radiation flux.  }
\label{fig::modelparams2}
\end{figure*}

Figs.~\ref{fig::modelparams} and \ref{fig::modelparams2} show
time-averages and snapshots of several quantities in the nonthermal
run. The snapshot in these comparisons (right side of each panel)
corresponds to $t=1.8 \times 10^5 \, t_g$ and the time-averaging (left
side of each panel) is done from $t=1.5\times10^4-2\times10^4 \, t_g$.

The top panel in Fig.~\ref{fig::modelparams} shows the gas density
$\rho$.  As expected, and as seen also in the control run
(Fig.~\ref{fig::compplots}), the disc is geometrically thick and
turbulent, the latter evident in the snapshot density distribution
(even more so in the temperature distribution discussed next). The
blue contour corresponds to the location where the accretion time-scale
$t_\text{acc}$ in the time-averaged model is equal to the
time-averaging duration $5000 \, t_g$. The accretion time scale is defined as
\begin{equation}
 \label{eq::tacc}
 t_\text{acc} \equiv \frac{r}{\sqrt{v_r^2 + r^2v_\theta^2}}.
\end{equation}
Since the total duration of the nonthermal run is $10^4\,t_g$, the
above limit is a conservative estimate of the region of inflow
equilibrium (it corresponds to the `strict' criterion, as defined in
\citealt{Narayan2012}). We can be confident that any region of the
flow that lies inside the surface defined by the above limit has
reached steady state and has forgotten the initial conditions when
relativistic electron injection was first turned on.

The second panel in Fig.~\ref{fig::modelparams} shows the electron
temperature, which ranges from $\sim10^{10}$\,K in the disc at $r\approx
30\, r_g$ to $10^{12}$\,K in the funnel region. In very low accretion
rate systems such as Sgr A*, both radiative cooling and Coulomb
coupling are weak and neither is capable of controlling the electron
temperature \citep{Yuan14}. The temperature is thus primarily
determined by the viscous heating and is highly dependent on the
heating fraction $\delta_e$. Our prescription \citep{Howes10} has high
$\delta_e\approx1$ in regions of high magnetization, which explains
the high temperature in the polar region (where $\beta<1$) compared to
the equatorial plane (where typically $\beta>5$).

The third panel in Fig.~\ref{fig::modelparams} shows the ratio of the
energies in nonthermal and thermal electrons. Since the radiative and
Coulomb coupling between the two species is weak, the energy ratio should
be set primarily by the injection ratio $\delta_\text{nth}$, which we
have fixed at 1.5 per cent throughout. In much of the equatorial plane out to
$r\approx30\,r_g$, the energy ratio is indeed approximately equal to
$\delta_\text{nth}$. Regions where the ratio is lower than
$\delta_\text{nth}$ correspond to places where the electron
temperature is lowest. In these regions, the overall electron heating
fraction $\delta_e$ is small and there has not been enough injection
of nonthermal particles to bring the energy up to the injection
value. Conversely, in the snapshot distribution, we see some regions
where the nonthermal-to-thermal energy ratio exceeds
$\delta_\text{nth}$. In these regions, the thermal electrons are
heated to high temperatures $\sim10^{12}$ K. At these temperatures,
the thermal electrons that produce most of the synchrotron emission
have Lorentz factors $\gamma > 500$, greater than the minimum
$\gamma_{\rm inj\,min}$ of the injected nonthermal electrons. These
high-$\gamma$ thermal electrons lose energy rapidly to radiation,
lowering their energy relatively more quickly compared to the
nonthermal electrons.

Finally, the fourth panel in Fig.~\ref{fig::modelparams} shows the
overall fraction of the electron population that is in the nonthermal
distribution. In the snapshot image, regions with a high ratio of
nonthermal electrons to the total population are coincident with
regions of high thermal electron temperature (second panel). This is
because both distributions are primarily driven by the fraction
$\delta_e$ of electron viscous heating (since $\delta_\text{nth}$ is
fixed).

In Fig.~\ref{fig::modelparams2}, we display quantities related to the
cooling and radiation from nonthermal particles. The top panel shows
the magnetic field strength, which is on average $\sim10$\,G
throughout much of the region in inflow equilibrium ($r \lesssim 40
\,r_g$). However, the snapshot on the right shows considerable
evidence for turbulence and deviations from the mean. Regions with a
stronger magnetic field in the snapshot image correlate with regions
of higher thermal electron temperature (second panel of
Fig.~\ref{fig::modelparams}); this is expected since the electron
energy injection fraction $\delta_e$ increases with magnetization. In
addition, since the nonthermal injection rate is proportional to
$\delta_e$, the same regions also stand out in the snapshot
distribution in the fourth panel of Fig.~\ref{fig::modelparams}.

The second panel in Fig.~\ref{fig::modelparams2} shows the magnitude
of the radiation flux $\hat{F}^i$, represented by the color scale,
with streamlines indicating the direction of the flux vector. Since
the accretion flow is highly optically thin, radiation is emitted
more-or-less isotropically and freely streams out of the system.

The third and fourth panels in Fig.~\ref{fig::modelparams2} show the
fluid frame power in radiation from thermal and nonthermal electrons,
respectively. The thermal emission dominates in the inner regions up
to $r \sim 10\,r_g$, and then declines rapidly at larger radii where
the electrons are cooler.  However, as previously discussed, highly
energetic nonthermal electrons are present even at large radii,
because of our simple injection prescription. Therefore, there is
significant nonthermal synchrotron emission out to $r \sim 50\,
r_g$. The two snapshot panels show that the instantaneous radiation
power in both thermal and nonthermal emission traces the regions of
strongest magnetic field in the top panel.

\subsection{Synchrotron Break}
\label{sec::synchbreak}

\begin{figure*}
\centering
\includegraphics*[width=\textwidth]{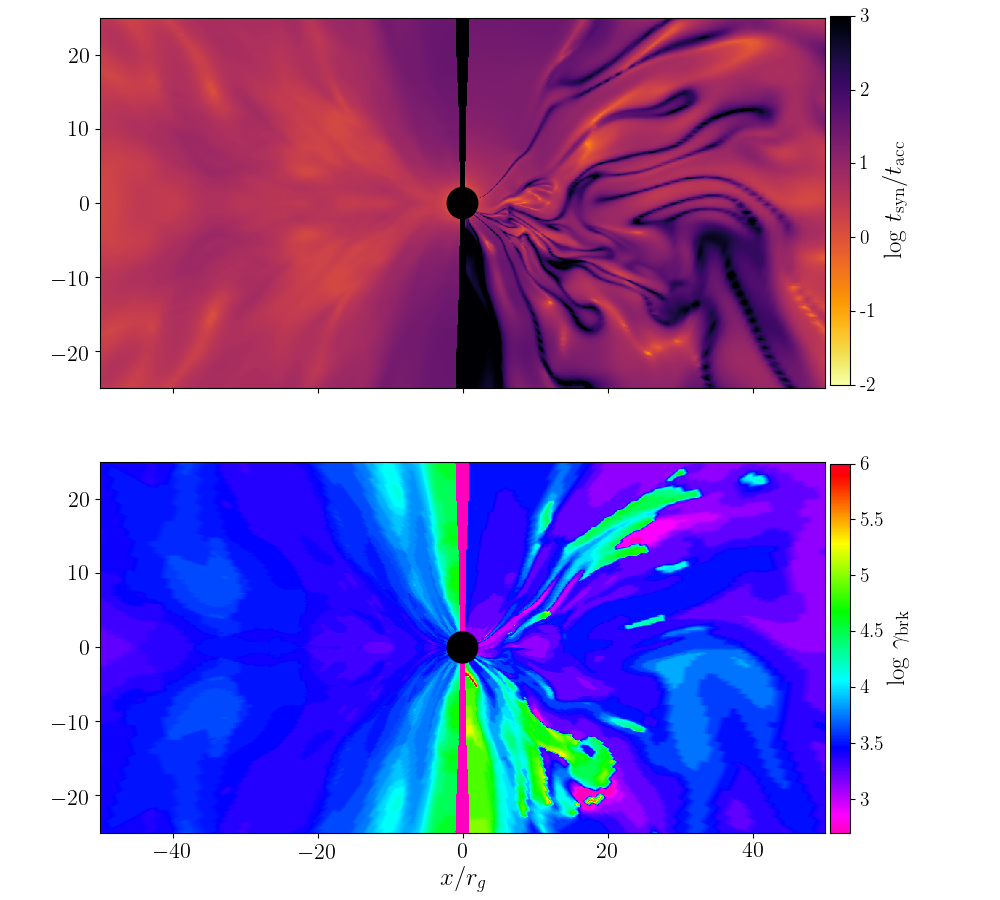} 
\caption{(Top) Ratio of synchrotron cooling time-scale to
  accretion time-scale, $t_\text{syn}/t_\text{acc}$, for a snapshot at
  $t=1.8\times10^4 \, t_g$ (right) and corresponding ratio computed from
  time-averaged primitives (left).  (Bottom) Location of the
  synchrotron cooling break Lorentz factor $\gamma_\text{brk}$.  The
  cooling break is at higher $\gamma$ in regions where
  $t_\text{syn}/t_\text{acc}$ is large. Electrons in such regions are
  advected away before they can be cooled by the magnetic field. }
\label{fig::timeratio}
\end{figure*}

The dominant physical processes shaping the evolution of the
nonthermal electron energy distribution $n(\gamma)$ in the nonthermal simulation
are electron injection and synchrotron cooling. As the nonthermal
particles cool via synchrotron emission, the spectrum will break from
the injection power-law slope $-p=-3.5$ to $-(p+1)=-4.5$. The
$\gamma_\text{brk}$ at which the break occurs moves to lower values
with increasing time. From equation~\eqref{eq::syncooltime}, under
constant injection and given a characteristic magnetic field strength
of $B\sim10$\,G, $\gamma_\text{brk}$ will move all the way down to
$\gamma_{\rm inj\,min}$ in $1.5\times10^4$ s, or $780 \, t_g$. However,
in the actual simulation, non-constant particle injection rates,
adiabatic compression, and advection modify the development of the
synchrotron break and can shift the break Lorentz factor to higher
$\gamma$, with advection having the strongest effect.

In the top panel of Fig.~\ref{fig::timeratio}, we plot the ratio of
the synchrotron cooling time (equation~\ref{eq::syncooltime}) to the
accretion/advection time-scale (equation~\ref{eq::tacc}). We find that
this ratio is $>1$ almost everywhere in the region considered, which
indicates that, before the spectrum can break fully, the
gas is advected away or falls into the black hole.

The second panel of Fig.~\ref{fig::timeratio} shows the Lorentz factor
$\gamma_\text{brk}$ of the synchrotron cooling break in the nonthermal
distribution. We determine $\gamma_\text{brk}$ in each cell simply by
finding the maximum of $\gamma^{p+1/2} n(\gamma)$. By the late times
we are considering, the cooling break has propagated to low Lorentz
factors, but since the accretion time-scale is shorter than the cooling
time-scale, the break still lies above
$\gamma_{\rm inj\,min}$. In much of the disc, the break is around
$\gamma_\text{brk}\sim3000$. In the funnel region, gas moves with high
velocities either into the BH or out along the axis; the corresponding
small inflow/outflow (advection) time-scale means that electrons do not
have enough time to cool before being swept away. Thus, the break
Lorentz factors in the funnel are typically higher than in the rest of
the simulation, $\gamma_\text{brk} \sim 10^4$.

In the time-averaged distribution, the ratio of synchrotron to advection times can
provide a quick estimate of the break Lorentz factor. In the funnel
regions, where $t_\text{syn}/t_\text{acc} \approx 100$, we can
estimate the position of the break by substituting $t_\text{syn}/100$
in equation~\eqref{eq::synbrk}; the result is $\gamma_\text{brk}
\approx 5\times10^4$. A comparison with the second panel shows that
this quick estimate is reasonably good.

The snapshot distribution of break Lorentz factor shows more structure
than the average. Much of this structure is due to the turbulent
magnetic field, which creates regions of short and long synchrotron
cooling times. However, the regions with
high $\gamma_\text{brk}$ do not always have a one-to-one
correspondence with regions of large $t_\text{syn}/t_\text{acc}$ (see
e.g. around $x=20 \, r_g$, $z=10 \, r_g$). This is because, in addition to
synchrotron cooling and advection, other processes -- particularly
adiabatic compression -- can shape the spectrum. Compression acts to
push the entire distribution to higher $\gamma$, so it naturally
pushes the break Lorentz factor to a higher $\gamma$ than predicted by
equation~\eqref{eq::synbrk}. In future studies, we will examine the
effects of adiabatic compression on the spectrum in more detail.

\subsection{Spectra and Images}
\label{sec::spectra}

\begin{figure*}
\centering
\includegraphics*[width=\textwidth]{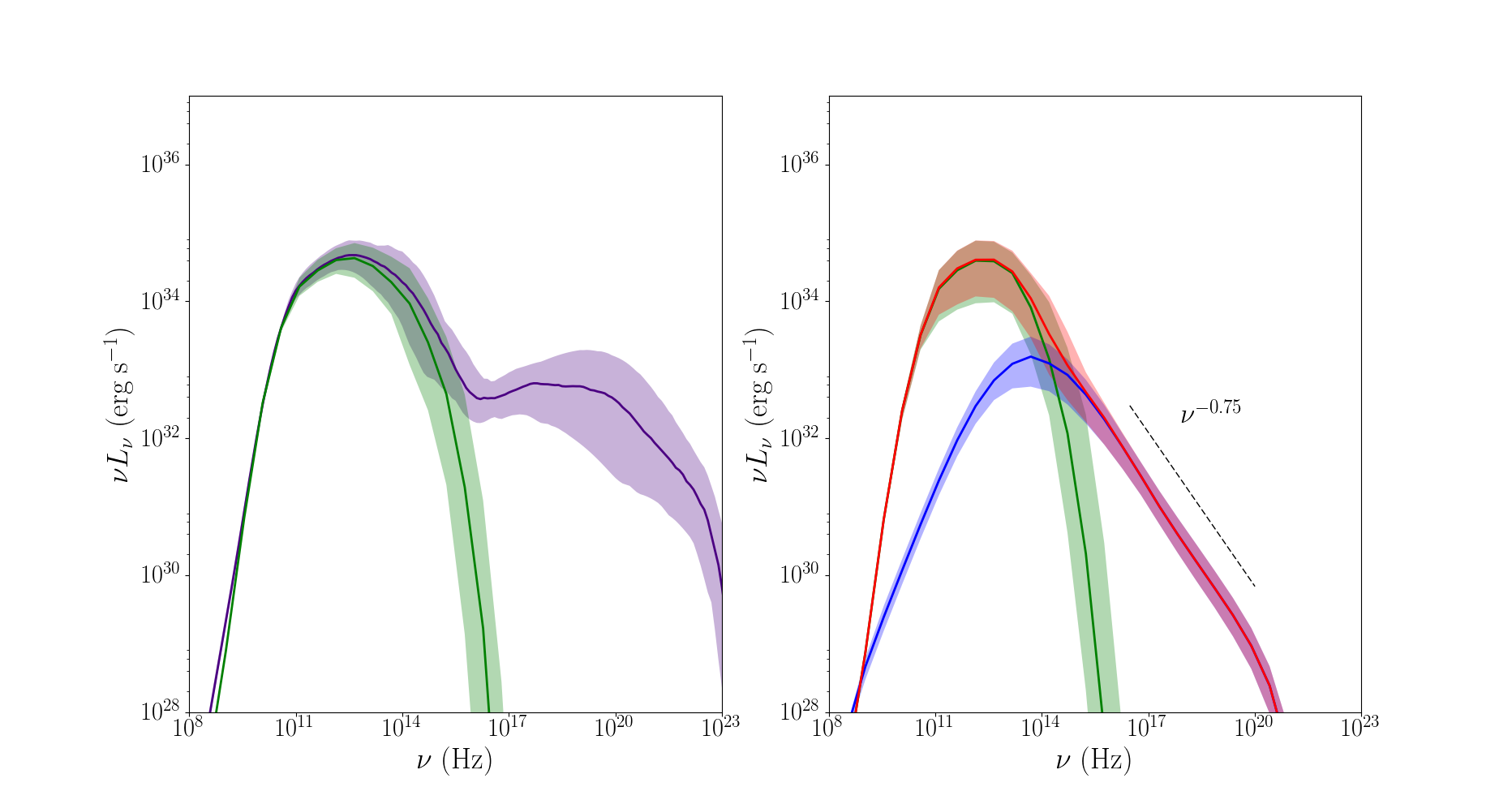} 
\caption{
  (Left Panel) Median spectral energy distribution (solid lines)
  of the thermal control run computed from the snapshot data from 15,000-20,000 $t_g$, 
  as observed at an angle of 60$^\circ$ with respect to the disc polar axis. The shaded regions
  represent the 68 per cent confidence interval (nominal 1$\sigma$ range) for the time-variability of
  the  spectra in this interval. 
  The green spectrum is obtained using \texttt{grtrans} \citep{Dexter16}, which includes
  only synchrotron radiation. The indigo spectrum was computed with
  \texttt{HEROIC} \citep{Narayan16_HEROIC}, including
  bremsstrahlung emission and inverse Compton scattering.  
  (Right Panel) Synchrotron-only spectra of snapshots from the nonthermal simulation in the
  range 15,000-20,000 $t_g$ computed with \texttt{grtrans}. The green and blue lines show the
  spectra of the thermal and nonthermal electrons, respectively, and
  the red line shows the total spectrum of both populations
  combined. The dashed line shows the expected power-law slope produced by the broken spectrum of
  nonthermal electrons, $L_\nu \propto \nu^{-p/2}\propto \nu^{-1.75}$.}
\label{fig::spectra}
\end{figure*}

\begin{figure*}
\centering
\begin{tabular}{lll}
\includegraphics*[width=.3\textwidth]{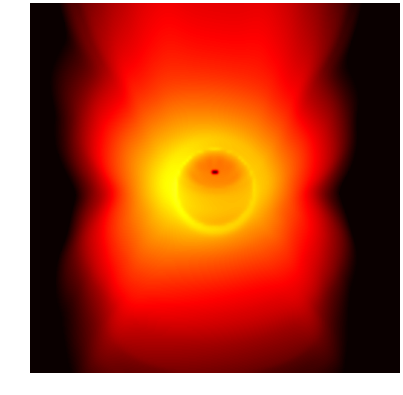} &
\includegraphics*[width=.3\textwidth]{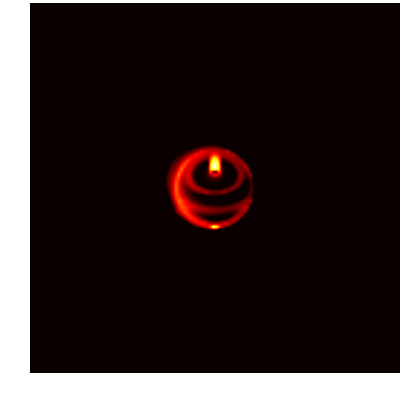} &
\includegraphics*[width=.3\textwidth]{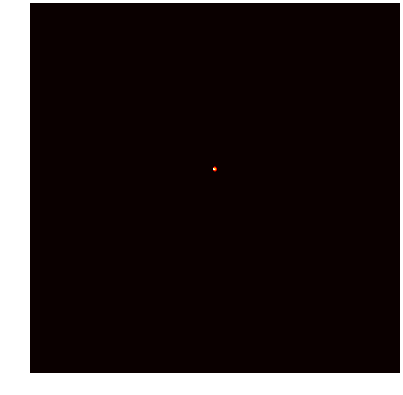} \\
\includegraphics*[width=.3\textwidth]{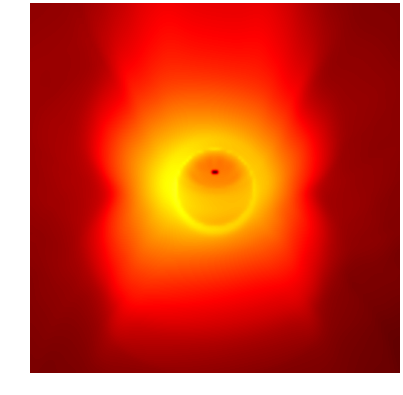} &
\includegraphics*[width=.3\textwidth]{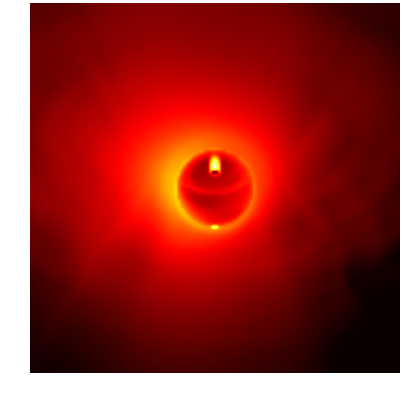} &
\includegraphics*[width=.3\textwidth]{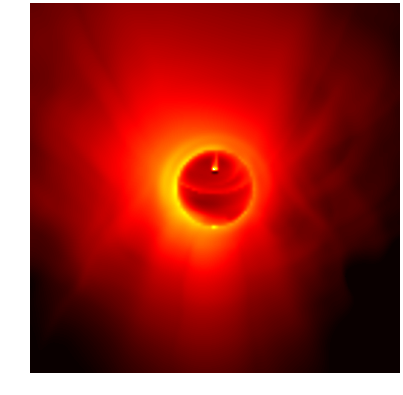} 
\end{tabular}
\caption{Images ($50 \, r_g$ wide, using logarithmic color maps) of synchrotron
  emission only, computed using \texttt{grtrans}, for the time-averaged
  control run (top row) and the nonthermal run (bottom row). The
  images correspond to 230\,GHz sub-millimeter emission (left), 136\,THz
  near-infrared emission (middle), and 2\,keV X-ray emission (right).}
\label{fig::images}
\end{figure*}

We computed spectral energy distributions (SEDs) for both the thermal-only control model and the
full nonthermal model using \texttt{grtrans},\footnote{https://github.com/jadexter/grtrans}
an open-source (polarized) ray tracing and radiative transfer code for black hole
spacetimes \citep{Dexter16}. We modified \texttt{grtrans} to compute
the nonthermal synchrotron emissivity $j_\nu$ and absorption
coefficient $\alpha_\nu$ directly from the local magnetic field and
the appropriate integrals over the nonthermal electron energy
distribution $n(\gamma)$ (\citealt{RL}, equations 6.33 and
6.50;\footnote{Note that to derive equation~\eqref{eq::alnu} from \citet{RL} 
equation 6.50, we perform an integration by parts and discard the boundary 
term.} for recent work on integrating polarimetric synchrotron emissivities 
for various electron distribution functions see \citealt{Leung11} and \citealt{Pandya16}). 
The integrals for $j_\nu$ and $\alpha_\nu$ are
\begin{align}
 \label{eq::jnu}
  j_\nu &= \frac{\sqrt{3}}{4\pi^2}\frac{e^3B \sin\alpha}{mc^2} \int n(\gamma) 
F\left(\frac{\nu}{\nu_c}\right) \rm{d}\gamma, \\
  \label{eq::alnu}
 \alpha_\nu &= \frac{4\pi}{3\sqrt{3}}\frac{e}{B\sin\alpha} \int 
\frac{n(\gamma)}{\gamma^5} K_{5/3}\left(\frac{\nu}{\nu_c}\right) \rm{d}\gamma.
\end{align}
In the above expression, $\alpha$ is the pitch angle between the line of sight 
and the magnetic field in the fluid frame, $F(x) = x\int_x^\infty K_{5/3}(y) dy$ 
is the synchrotron function, and $\nu_c$ is  the characteristic synchrotron 
frequency,
\begin{equation}
 \nu_c = \frac{3\,eB\,\gamma^2 \sin\alpha }{4\pi m_e c}.
\end{equation}

We performed the radiative transfer only for the total intensity (we
plan to include polarization in the future). To speed up the
computations, we used fitting functions for the synchrotron function
$F(x)$ and Bessel function $K_{5/3}(x)$ from \citet{Fouka}.

The green curve in the left panel in Fig.~\ref{fig::spectra}
shows the median \texttt{grtrans} synchrotron SED from the thermal control run which was
computed from the snapshot data from 15,000-20,000 $t_g$, 
as observed at an angle of 60$^\circ$ with respect to the disc polar axis. 
The shaded region represents the 68 per cent confidence interval (nominal 1$\sigma$ range) 
for the time-variability of the spectrum in this interval. The spectrum peaks at
$\nu\sim 10^{12}$\,Hz, with a steep fall-off at lower frequencies
because of self-absorption and a fall-off at higher frequencies
because of the rapid decline in the number of thermal electrons at
larger Lorentz factors. 


The indigo curve in the same panel was computed using the
post-processing code \texttt{HEROIC}, \citep{Narayan16_HEROIC, Zhu_2015}, which 
self-consistently solves for the spectrum and
angular distribution of radiation at each position using the
radiative transfer equation. The \texttt{HEROIC} radiative transfer includes all
radiation processes --- synchrotron, bremsstrahlung, and inverse
Compton scattering. In the synchrotron
component, the \texttt{HEROIC} spectrum agrees very well with the
\texttt{grtrans} spectrum except at frequencies below
$10^{10}$\,Hz. This small discrepancy arises because the \texttt{HEROIC}
computations are done using simulation data out to a radius of
$300\,r_g$, whereas the \texttt{grtrans} calculations are limited to
$50\,r_g$, so \texttt{HEROIC} picks up more low-frequency emission from further out
in the simulation volume. While the power in nonthermal radiation extends to larger radii
than the thermal population (Fig.~\ref{fig::compplots}), nonthermal spectra from \texttt{grtrans} generated using data out to $100\,r_g$ 
are nearly identical to the plots shown using a maximum radius of $50\,r_g$. 


The right panel in Fig.~\ref{fig::spectra} shows spectra of the nonthermal run, 
computed with \texttt{grtrans} using the same parameters as the  left panel.  
Similarly to the left panel, the solid lines are the median SEDs from the interval 15,000-20,000 $t_g$, 
and the shaded regions are the 68 per cent confidence range of the time variability.
Since \texttt{HEROIC} does not presently include
nonthermal electrons, we do not show comparison spectra
from that code. Comparing the thermal-only (green curve) and the
nonthermal-only (blue curve) \texttt{grtrans} spectra, we see that
thermal emission dominates by far in the sub-millimeter band,
nonthermal emission is modestly stronger at infrared wavelengths, and
is the only contributor to the synchrotron emission at X-ray wavelengths. 
The power-law synchrotron emission is optically thin, and shows a characteristic
slope $L_\nu \propto \nu^{1/3}$ at low frequencies. 
The power-law tail in the spectrum at high frequencies has
a spectral slope $L_\nu \propto \nu^{-p/2} \propto \nu^{-1.75}$, as expected for a
population of electrons with a distribution mostly broken to a power-law slope $-(p+1)=-4.5$.

The red curve in Fig.~\ref{fig::spectra} shows the combined
synchrotron emission from both thermal and nonthermal electrons. By
and large, the combined spectrum is a direct sum of the two
independent contributions, except at the lowest frequencies,
where absorption by thermal electrons suppresses the nonthermal
emission \citep{Ozel2000, Yuan2003}. This effect is seen also
in other recent studies in which synchrotron spectra from thermal and
nonthermal electrons are computed by post-processing single
temperature GRMHD simulations \citep{Ball_16, Mao2016}. Note that the spectra shown here include only thermal
and nonthermal synchrotron emission. For more realistic nonthermal
spectra, it will be necessary to incorporate synchrotron,
bremsstrahlung, and inverse Compton scattering from nonthermal
electrons into a global radiative transfer solver like \texttt{HEROIC}
or a Monte Carlo transfer code such as \texttt{grmonty}
\citep{grmonty}.

Sgr A* is known to be more variable in the infrared compared to
sub-millimeter, and even more variable in X-rays \citep{Eckart06, Yusef06, 
Dodds09, Neilsen}. From the variability in the spectra shown in the right panel of
Fig.~\ref{fig::spectra}, it is clear that our uniform injection prescription generates 
little variability in the nonthermal synchrotron emission at high
frequencies. However, the present simulations are not suitable for exploring the variability in detail,
both because they are in 2D ---\citet{Sadowski_3D} 
show that the variability properties of 2D and 3D
simulations are different --- and because we have used a toy
prescription for nonthermal energy injection. The thermal spectrum in Fig.~\ref{fig::spectra} 
shows that variability in the thermal X-ray inverse Compton
spectrum exceeds that in the direct synchrotron emission at lower frequencies. Furthermore, a direct comparison
of the thermal and nonthermal frequency-integrated inverse Compton power shows that while the thermal IC power dominates
in the disc in the densest regions at small radii, the high energy of the nonthermal electrons (and the fact that the IC power grows
as $\gamma^2$) leads to the nonthermal IC power exceeding the thermal IC power in the funnel region and 
in the disc at radii $\gtrsim 40 \, r_g$. Thus, we expect the nonthermal electrons to make a significant contribution to the high
frequency spectrum and variability from IC emission. In sum, to accurately explore variability and flares from nonthermal electrons, we will need to 
extend our simulations to 3D, implement local injection prescriptions, and include bremsstrahlung
and inverse Compton emission in the nonthermal radiative transfer. This is a promising
direction for future work.

Fig.~\ref{fig::images} shows \texttt{grtrans}-generated ray-traced
images of the synchrotron emission from the time-averaged simulations at 3 
frequencies: 230\,GHz, which is near the thermal synchrotron peak and corresponds to the
observing frequency of the Event Horizon Telescope \citep{Doeleman08},
136\,THz in the near infrared, and $4.8\times10^{17}$\,Hz (2\,keV) in
X-rays. The images are $50$ projected gravitational radii across and displayed in a log scale.
The bright regions of the image at 230\,GHz are practically
the same for the thermal and nonthermal runs, confirming that much of
the emission is from thermal electrons. There is, however, additional
extended flux at large radii in the bottom panel because of emission
by nonthermal electrons. The ring in the infrared image is brighter
when nonthermal electrons are included, and the emission extends to
noticeably larger radii. The X-ray image is almost entirely from
nonthermal emission. As for the spectra in Fig.~\ref{fig::spectra}, these results depend
sensitively on the simple nonthermal energy injection prescription we have
used. Bremsstrahlung emission and inverse Compton scattering will also modify these
images, especially in X-rays

\section{Summary and Conclusions}
\label{sec::summary}

In this paper, we introduced a new algorithm to self-consistently
evolve a population of nonthermal electrons in a black hole spacetime,
in parallel with magnetized thermal gas and radiation.  In each time
step, a fraction of the viscously generated heat is used to heat
some of the thermal electrons and to transfer them to the nonthermal
population. The nonthermal electrons move with the fluid, and their
energy distribution is modified by gas compression and expansion,
Coulomb coupling, and radiative cooling. The back-reaction of the
nonthermal electrons on the thermal population is automatically
included.

We validated the algorithm on a variety of test problems, and
presented first results on a 2D black hole accretion flow with
nonthermal electrons. This simulation has a low mass accretion rate,
roughly equal to the rate estimated in \sgraa. As a result, the
nonthermal distribution does not significantly affect the gas dynamics
or thermodynamics of the thermal electrons or ions. However, the
radiation power is enhanced, since the nonthermal electrons radiate
more efficiently than their thermal counterparts. Furthermore, the
energy distribution of the nonthermal electrons varies with location
in the accretion flow. The distribution exhibits a synchrotron cooling
break, and the break Lorentz factor $\gamma_\text{brk}$ varies with
position, being set by local conditions such as the magnetic field
strength (which determines synchrotron power) and the gas velocity
(which sets the effective advection time); $\gamma_\text{brk}$ is also
modified by other factors such as strong adiabatic compression.

The current work considers only one particularly simple prescription
for injection into the nonthermal population. Specifically, we inject
a fixed fraction of the local electron viscous heating into
the nonthermal distribution, and we inject the nonthermal particles
over a fixed range of $\gamma$, with a fixed power-law slope.  The
resulting simulation results are strongly influenced by these choices.
A constant injection range of $\gamma$, independent of radius, ensures
that nonthermal synchrotron emission dominates over thermal emission
at large radii, where the temperature of the thermal electrons falls
off rapidly. This is reflected in Fig.~\ref{fig::images}, which shows
that at high frequencies, nonthermal electrons from farther out in the
disc dominate the raytraced synchrotron image of the accreting gas.
Furthermore, the choice of a minimum injection Lorentz factor
$\gamma_{\rm{inj\,,min}}=500$ means that most of the nonthermal
emission is concentrated at infrared or higher frequencies, while the
image at 230 GHz, of interest for the Event Horizon Telescope, iss
basically unchanged compared to a purely thermal model. In principle,
$\gamma_{\rm{inj,min}}$ should be chosen such that the nonthermal
population connects smoothly to the thermal distribution, without a
gap between the two. This will be necessary when we use the code to
model real systems such as \sgraa.

Another consequence of our choice of injection parameters is that the
high frequency nonthermal emission in our simulation shows relatively
little time variability. This is because nonthermal electrons are
distributed relatively smoothly and uniformly throughout the
simulation. The rapid variability that is observed in \sgra is likely
driven by strong localized injection, perhaps from shocks or magnetic
reconnection. This suggests a much more sporadic and localized
injection of nonthermal energy, with small regions where the fraction
of energy going into the nonthermal electrons, $\delta_\text{nth}$, is
much larger than the $1.5$ per cent we chose in this work, and large regions
elsewhere with $\delta_\text{nth}$ near-zero (see \citealt{Ball_16}). 
Furthermore, particle-in-cell simulations show
that electrons accelerated in reconnection events attain progressively
harder energy spectra as the magnetization of the plasma increases
\citep{Sironi14}. This will again have a strong impact on
variability. We plan to investigate this issue in a future study.

Finally, we note that many of the results presented here are specific
to accretion at ultra-low mass accretion rates. At higher accretion
rates, the radiative efficiency will be larger, as will the scattering
optical depth, which will increase the importance of inverse Compton
scattering. Klein-Nishina corrections, which are included in the code
but had a minimal effect in the present study, may become more
important for the nonthermal electrons. Also, while feedback from
nonthermal electrons on the other gas quantities was found to be
negligible in this work, it is likely to have more of an effect at
higher accretion rates, or with different injection
prescriptions. Two-temperature simulations with thermal electrons and
ions show that, at higher accretion rates, radiation from thermal
electrons cools the gas more effectively and causes the disc to be
thinner \citep{KORAL16, SadowskiGaspari}.  This
effect is likely to be enhanced when nonthermal electrons are included.

\section*{Acknowledgements}
We thank Jason Dexter for his help in using and modifying
\texttt{grtrans} and the referee, Sean Ressler, for his helpful comments. 
AC was supported in part by NSF grant AST-1440254.
RN was supported in part by NSF grant AST-1312651. AS
acknowledges support for this work by NASA through Einstein
Postdoctoral Fellowship number PF4-150126 awarded by the Chandra X-ray
Center, which is operated by the Smithsonian Astrophysical Observatory
for NASA under contract NAS8-03060.  The authors acknowledge
computational support from NSF via XSEDE resources (grant
TG-AST080026N) and from the PL-Grid Infrastructure.

\bibliography{ElecEv}

\end{document}